\newif\ifSC
\SCtrue
\ifSC
\documentclass[onecolumn,draftclsnofoot,12pt]{IEEEtran}
\else
\documentclass[10pt,twocolumn]{IEEEtran}
\fi

\usepackage{cite}
\usepackage{amsfonts}
\usepackage{bbm}
\usepackage{color}
\usepackage{amsmath}
\usepackage{amsthm}
\newtheorem{theorem}{Theorem}
\newtheorem{pf}{Proof}
\newtheorem{coro}{Corollary}[theorem]
\newtheorem{remark}{Remark}

\newcommand{\ie}{{\em i.e. }}
\newcommand{\ud}{u_{\mathbf{x}_\mathrm{d}}}
\renewcommand{\ud}{u_{\rd}}
\newcommand{\rd}{r_{\mathrm{d}}}

\newcommand{\sx}{s_{\mathbf{x}}}
\newcommand{\tbnproc}{\Phi}
\newcommand{\phim}{\Phi_\mathrm{M}}
\newcommand{\phitotal}{\Phi_\mathrm{T}}
\newcommand{\x}{\mathbf{x}}
\newcommand{\xd}{\mathbf{x}_\mathrm{d}}
\newcommand{\xt}{\mathbf{x}_\mathrm{t}}
\newcommand{\1}{\mathbbm{1}}
\newcommand{\norm}[1]{\| #1\|}

\newcommand{\erfc}[1]{\mathrm{erfc}\left\{#1\right\}}
\newcommand{\exps}{\mathbb{E}}
\newcommand{\ts}{T_{\mathrm{S}}}
\newcommand{\Es}{\mathrm{E_S}}
\newcommand{\Ei}{\mathrm{E_I}}
\newcommand{\Em}{\mathrm{E_C}}
\newcommand{\Et}{\mathrm{E_T}}
\newcommand{\ths}{\mathrm{th}}
\newcommand{\expS}[1]{\exp\left(#1\right)}
\newcommand{\expU}[1]{e^{#1}}
\newcommand{\prob}[1]{\mathbb{P}\left[#1\right]}
\newcommand{\expect}[1]{\mathbb{E}\left[#1\right]}
\newcommand{\dd}{\mathrm{d}}
\newcommand{\pzero}{P_{\mathrm{0}}}

\newcommand{\pone}{P_{\mathrm{1}}}
\newcommand{\tbl}[1]{\tau_{\mathrm{L}#1 }    }

\newcommand{\tbh}[1]{\tau_{\mathrm{H}#1 }    }
\newcommand{\dfracnrho}[2]{\frac{\partial^{#1} #2}{\partial \rho^{#1} }}

\renewcommand{\pzero}{p_0}
\renewcommand{\pone}{p_1}
\newcommand{\probe}{\mathsf{P}_\mathrm{e}}
\newcommand{\probeb}[1]{\mathsf{P}_\mathrm{e#1}}
\newcommand{\dist}[2]{g_{\mathsf{R}_\mathrm{d}}\left(#2\right)}
\newcommand{\efunc}[3]{e_{#1,#2}(#3)}
\newcommand{\opt}[1]{#1_{\mathrm{opt}}}
\newcommand{\cb}{{\mathsf{b}}}

\newcommand{\resultheading}[1]{\vspace{-.09in} \ \\ \textit{\textbf{#1}}:}

\newcommand{\expects}[2]{\mathbb{E}_{#1}\left[#2\right]}

\newcommand{\w}{w}
\newcommand{\wC}{W}
\newcommand{\indside}[1]{\1\left(#1\right)}
\newcommand{\noisiqfunc}{P}
\newcommand{\isiqfunc}{Q}
\newcommand{\cir}[2]{f(#1,#2)}
\newcommand{\cirslot}[2]{h_{#1}[#2]}

\ifCLASSINFOpdf
\usepackage[pdftex]{graphicx}
\DeclareGraphicsExtensions{.pdf}
\else
\fi

\title{Analysis of  Diffusion Based Molecular Communication   with Multiple Transmitters having Individual Random Information Bits}

\begin{document}

\author{
Nithin V. Sabu and Abhishek K. Gupta
\thanks{The authors are with the department of Electrical Engineering, Indian Institute of Technology Kanpur, Kanpur, India 208016. (Email: nithinvs@iitk.ac.in and gkrabhi@iitk.ac.in). This research was supported by the  Science and Engineering Research Board  (DST, India) under the grant SRG/2019/001459.}}

\maketitle

\begin{abstract}
In this paper, we present an analytical framework to derive the performance of a molecular communication system where a transmitter bio-nano-machine (TBN) is communicating with a fully-absorbing spherical receiver bio-nano-machine (RBN) in a diffusive propagation medium in the presence of other TBNs. We assume that transmit bits  at each TBN is random and different than transmit bits at other TBNs. We model the TBNs using a marked Poisson point process (PPP) with their locations as points of PPP and transmit symbols as marks. We consider both inter-symbol interference (ISI) and co-channel interference (CCI). ISI is caused by molecules transmitted in the previous slots while CCI is due to the molecules emitted from other TBNs. We derive the bit error probability of this system by averaging over the distribution of the transmit bits as opposed to the past approaches consisting of conditioning on previous transmit bits and/or assuming the transmit bits of every TBN are the same. Using numerical results, we validate our analysis and provide various design insights about the system, for example, the impact of detection threshold on the system performance. We also show the importance of accurately incorporating the randomness of transmit bits in the analysis.
	
\end{abstract}


\section{Introduction}
Molecular communication is a communication paradigm inspired from the nature that includes
communication between macro-scale, micro-scale and nano-scale devices (or organisms) with the help of molecules as information carriers between these devices.  An example of molecular communication is the human body itself where most communications including intra-cellular, inter-cellular, and inter-organ communications occur via various types of molecules \cite{suda2005exploratory,hiyama2006molecular}. Molecular communication can enable nano-machines (devices with nano-scale functional units) to communicate over small distances (typically several micrometers) in an appropriate medium.
%
%
Nano-machines can be biological systems like bacteria, human cells, which can perform simple computations, sensing and actuation or artificially created devices to mimic such activities. These nano-machines acting as transmitters or receivers can communicate with each other by sending and receiving messenger  molecules. These molecules are termed as \textit{information molecules} (IMs). 
The molecular communication using bio-nano-machines (nano-machines made up of biological materials) may consist of five basic steps- namely- encoding, transmission, propagation, receiving and decoding \cite{suda2005exploratory,nakano2013molecular,nakano2011biological}. First, the transmitter bio-nano-machine (TBN) encodes the transmit message  into IMs via various schemes {\em e.g.} by using different concentration or types of IMs for each message, or by emitting IMs at different time instants, or by encoding in the three-dimensional structure of molecules \cite{farsad2016comprehensive}. 
Then, the transmission step consists of emitting these IMs to the propagation medium  via various mechanisms including  budding of vesicles, or opening gate channels in the membrane. These emitted IMs, then, move from  TBN to the receiver bio-nano-machine (RBN). This propagation can be either controlled {\em e.g.} via movement of motor protein over molecule rails  or can be passive {\em e.g.} diffusion via Brownian motion \cite{einstein1956investigations}. 
At RBN, IMs are captured  
using receptor structures which can bind to IMs. In the decoding phase, the captured molecules are used to estimate the transmitted message. 
Owing to its bio-compatibility, energy efficiency and high storage capacity,  molecular communication has many futuristic applications including nano-machine communication, molecular computing, targeted drug delivery  \cite{nakano2013molecular} and is seeing a growing interest among researchers.

\textit{\textbf{Related Work:}} In molecular communication via diffusion (MCvD), IMs travel in the  medium by Brownian motion. A physical end-to-end model suitable for the study of molecular communication was demonstrated in \cite{pierobon2010physical}. The signal strength of point-to-point molecular communication channel was studied in \cite{kuran2010energy,yilmaz2014simulation}. The channel characteristics for a 3D MCvD system with absorbing receiver was derived in \cite{yilmaz2014three}. 
In a MCvD system, there can be  multiple TBNs randomly distributed in the propagation medium which may also emit IMs of the same type. The emitted IMs from these interfering transmitters also reach the receiver  to cause interference termed as multi-transmitter interference or  co-channel interference (CCI) \cite{pierobon2012intersymbol,Jiang2014}. The CCI effects based on spatial distribution and characteristics of two TBN-RBN couples were evaluated in \cite{MehmetSukruKuran2011}. \par
Stochastic geometry \cite{AndDhiGup2016} has emerged as a tractable tool to study communication systems with random geometry and has been used to analyze the performance of MCvD systems owing to the random nature of TBNs and RBNs locations in the medium. The position of the bio-nano-machines in 3D spaces can be modeled using Poisson point process (PPP). For example, the spatial distribution of bacterial colonies inside cheese was shown to fit a PPP \cite{Jeanson1493}. An interference model  when the transmitter nano-machines are spatially distributed as uniform PPP, was presented in \cite{pierobon2014statistical} and the  probability distribution of the power spectral density of the received signal was derived. 
The work \cite{deng2017analyzing} presented a general model for collective signal strength at the passive and fully absorbing spherical receivers in a large-scale system where  transmitters are distributed according to a PPP 
and all transmitters are transmitting the same bits. Using stochastic geometry, the authors have   derived the bit error probability for the same. 
Due to their Brownian motion, molecules emitted by the TBNs at a time slot can travel in the propagation medium for a long time and can arrive at the RBN in later time slots. This results in inter-symbol interference (ISI) at the RBN \cite{pierobon2012intersymbol,kuran2010energy}. The receiver cannot distinguish between the desired and the ISI molecules. The effect of ISI on collective signal strength was not discussed in \cite{deng2017analyzing}.
The work \cite{dinc2017theoretical} considered both ISI and CCI, and derived the expected number of molecules  received at a fully absorbing receiver for a system where 
 the number of interfering transmitters is constant. 
The work \cite{dissanayake2018enhancing} considered a MCvD system where the interfering transmitters are  distributed as a homogeneous PPP and derived the analytical expression for the collective signal strength  and the bit error probability at the partially absorbing spherical receiver.\par 
%
In all the past works \cite{deng2017analyzing,dissanayake2018enhancing,dissanayake2019interference}, the probability of bit error is calculated by conditioning on the current and previous symbols of all the transmitters, and/or assuming that all the transmitters are sending the same bit sequence.
When the transmitters location are distributed as PPP and they are sending different transmit bits to their receiver counterparts, it is not possible to derive bit error probability by conditioning on the transmit bits of other transmitters. This is due to the fact that the number of transmitters is itself random and averaging over PPP cannot be performed without first averaging over transmit bits of transmitters.
In a system, where each TBN has its individual
information content (which can be distributed according to an arbitrary probability distribution over information symbols and different than the information content of other TBNs) that need to be sent to its RBN counterparts, it is very crucial to include its impact in the system performance by properly averaging over probability distribution of current and previous information bits of other transmitters.  The information content's randomness, independence across TBNs and its impact on the system's performance  was not studied in the past which is the one of the focus of this work.

\textit{\textbf{Contributions:} }  In this work, we consider a MCvD system with multiple TBNs in a three dimensional (3D) space.  We assume that the transmit bits (or information symbols) at each TBN is random and independent of the transmit bits at other TBNs.  All TBNs are using on-off keying (OOK) modulation scheme with the same  type of molecules for communication. We consider a typical fully absorbing receiver (RBN) at the origin. The distance of its associated TBN from the typical RBN is assumed to be a random variable (with fixed distance as a special case).  Due to multiple TBNs communicating in the medium,  both ISI and CCI would be present.   
Unlike the previous works  \cite{deng2017analyzing,dissanayake2018enhancing,dissanayake2019interference}, we derive the expected number of information molecules observed at the RBN and the probability of bit error by considering randomness of information bits and independence of data transmitted by TBNs. Note that the number of these bits is random as it depends on the number of transmitters.  At first the  error probability need to be averaged  over the previous bits of each transmitter, and then need to be averaged over the locations of transmitters using the point process distribution. Hence, the bit error probability cannot be simply obtained by  averaging the conditioned bit error probability derived in the past works. To correctly model the effect of these current and previous bits of all TBNs, marked version of PPP can be used where we model the interfering TBNs as a marked PPP with their location as points of the point process (PP) and deliverable information symbols as marks.  Modeling using marked PPP allows us to include randomness and independence of transmit bits in the analysis and perform the appropriate averaging. The analysis of the proposed system requires novel framework and derivation techniques compared to the existing literature. In this paper, we  derive the performance of this system when the previous bits at the desired TBN and the current and previous bits at the interfering TBNs are random.We model the interfering TBNs as a marked PPP with their location as points of PP and deliverable information symbols as marks.
  %
  In particular, the contributions of this paper are as follows:
  \begin{enumerate}
  	\item We provide an analytical framework for a molecular communication system consisting of multiple TBNs, each having  random information content (transmit symbols). We assume that the transmit symbols at each TBN are distributed according to  probability distribution and are independent of other TBNs.
  	\item We consider a typical fully-absorbing spherical receiver at the origin and derive the expected number of  desired and interfering molecules while considering molecular degradation, ISI and CCI.
  	\item We first derive the probability of bit error for system with no-ISI. 
  	We then extend the analysis to  systems with both ISI and CCI to derive the probability of bit error. The performed analysis helps us to understand the impact of various system parameters including detection threshold, TBN's density, molecular degradation, symbol time and evaluate their optimal values.
  	\item  We also provide insights about the considered system via numerical results. We highlight that the detection threshold plays a crucial role in the feasibility of molecular communication and should be adapted according to various propagation conditions {\em e.g.} transmitter-receiver distance. We also show the importance of accurately incorporating the randomness of transmit bits in the analysis.
  \end{enumerate}

\section{System model}

\begin{figure}
	\centering
	\includegraphics[width=.4\linewidth]{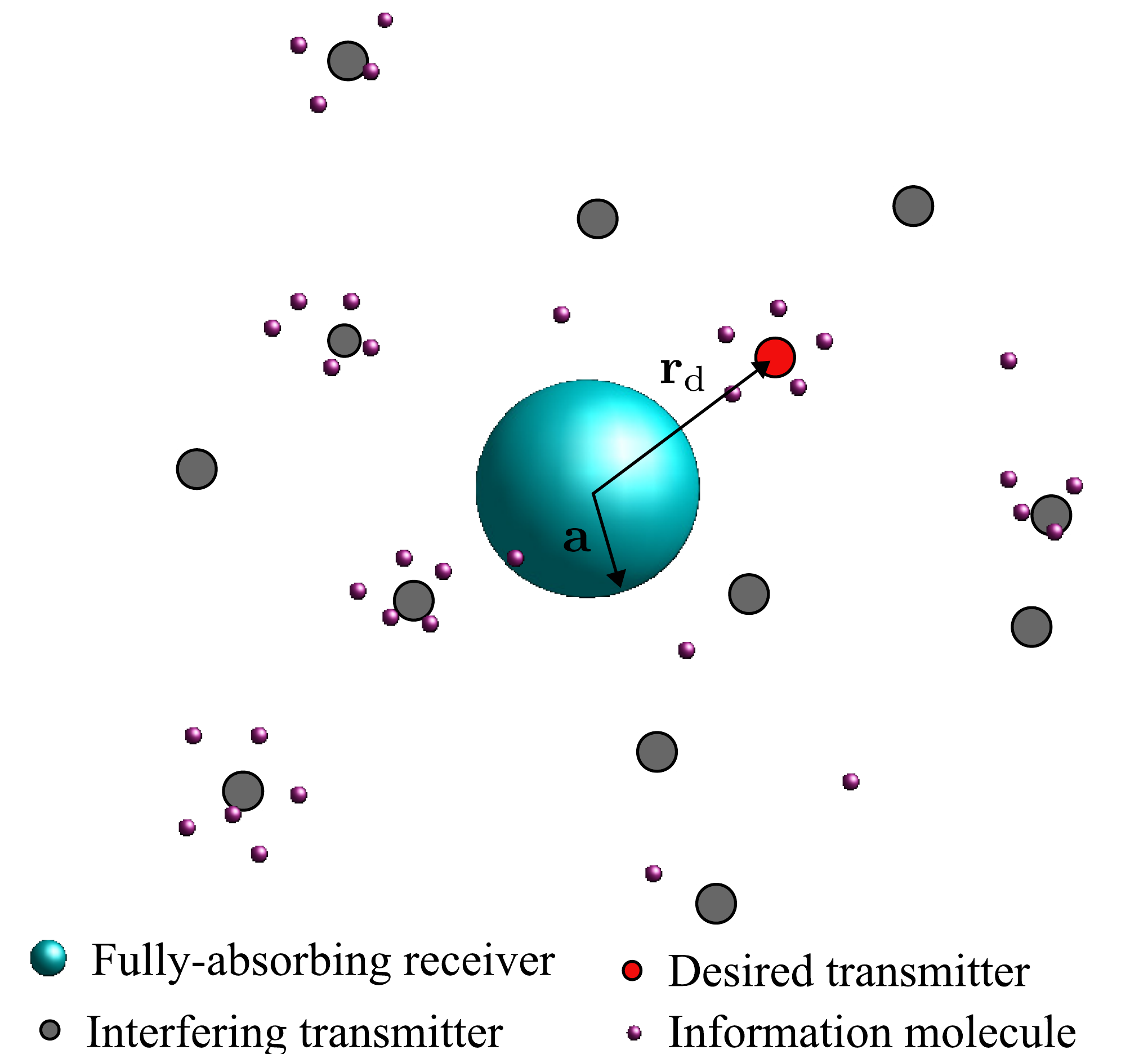}
	\caption{System model. A molecular communication system  with  a typical spherical fully absorbing RBN at the origin. The desired TBN (shown as red circle) is at a the distance  $\rd$ (which can either be fixed or a random variable
		).
	 The interfering TBNs (shown as grey circle) form a MPP.}
	\label{sm}
\end{figure}
\begin{table}[htbp]
	\caption{Notation Summary}
	\begin{center}
		\begin{tabular}{|p{.7in}|p{3.4in}|}
			\hline
			\textbf{Symbol}&\textbf{Definition} \\
			\hline
			$D$& Diffusion coefficient of the IMs in the propagation medium.\\
			\hline
			$\mu$& Molecular degradation rate.\\
			\hline
			$\ts$& Symbol duration.\\
			\hline
			$\cb$& Transmitted bit at the current slot of tagged TBN.\\
			\hline
			$\probeb{\cb}$& Probability of incorrect decoding of bit $\cb$.\\
			\hline
			$\pzero,\pone$& Probability of transmitting bit 0 and 1 respectively.\\
			\hline
			$\probe$& Total probability of bit error.\\
			\hline
			$\phim, \lambda$& Marked point process modeling TBN in $\mathbb{R}^3$ with density $\lambda$. \\
			\hline
			$\norm{\x}$& Distance of the point $\x$ from the origin.\\
			\hline
			$\xd,\ \rd$ & $\xd$ is the location of the tagged transmitter and $\rd=\norm{\xd}$.\\
			\hline
			$\phitotal$&  PP denoting union of $\phim$ and the tagged TBN at $\xd$ \ie $\{\xd\} \cup \phim$. \\
			\hline
			$\xt$ & Location of points in $\Phi_{\mathrm{T}}$.\\
			\hline
			$\cirslot{\norm{\x}}{l}$ &Fraction of IMs reaching the RBN in $l\ths$ slot  (\ie in the time window $[l\ts ,(l+1)\ts ]$) since the transmission.\\
			\hline
			$u_\x[0],u_\x[-1], $ $\cdots ,u_\x[-(L-1)]$& Number of molecules emitted by the transmitter located at distance $\norm{\x}$ from the typical RBN in the current time slot and past $L-1$ time slots respectively.\\
			\hline
			$u_\x[k_1:k_2]$& Vector denoting the number of molecules emitted in slots $k_1,k_1+1 \cdots k_2$ by the transmitter located at distance $\norm{\x}$.\\
			\hline
			$y\sim\mathcal{P}(\nu)$ & $y$ is Poisson distributed with parameter $\nu$.\\
			\hline
			$\mathfrak{B}_{n}(a_1,\cdots,a_n) $& The $n^{\text{th}}$ complete exponential Bell polynomial \cite{comtet2012advanced}.\\           
			\hline
			
		\end{tabular}
		\label{tab1}
	\end{center}
\end{table}
In this paper, we consider a molecular communication system in a 3D fluid medium without flow, as shown in Fig. \ref{sm}. The important symbols and notations are given in Table \ref{tab1}. The system consists of multiple TBNs. Each TBN is assumed to be a point source, which emit IMs to the propagation medium based on OOK modulation. 

\subsection{Network Model}
We consider a typical receiver at the origin which is a spherical and fully-absorbing receiver of radius $a$. However, the developed analytical framework can be used for a general receiver. This receiver consists of receptors at its surface, which can sense only single type of molecules. 
All molecules reaching the receiver surface are absorbed and counted for demodulating the transmitted information.

The TBN associated with the typical RBN, termed \textit{tagged} TBN, is assumed to be located at $\xd$. The distance $\rd=\norm{\xd}$ from the typical RBN can be a constant or a random variable. In addition to the tagged TBN, there are interfering transmitters in 3D fluid medium whose locations can be modeled by 3D homogeneous PPP $\tbnproc$ \cite{deng2017analyzing,dissanayake2019interference,deng20163d}. Since the
receiver occupies the space $\mathcal{B}(0,a)$,  the support of PPP is taken as $\mathbb{R}^3\setminus\mathcal{B}(0,a)$  \cite{deng2017analyzing,deng20163d}.
Let 
$\phitotal$ denote $\{\xd\}\cup \tbnproc$ \ie the union of the location of the desired transmitter and the TBNs PPP, $\Phi$.

\subsection{Modulation and transmission model}
Let $s_{\x}[i]$ denote the transmit bit of the point TBN located at $\x$ for the $i\ths$ time slot (of duration $\ts$).  $s_{\x}[i]$  is assumed to be a Bernoulli RV with parameter $\pone$.  At the beginning of the $i\ths$ time slot,  this point TBN emits $u_\x[i]$ number of molecules in to the propagation medium. $u_\x[i]$ can be either zero or $N$  according to the transmit bit ($s_{\x}[i]$). Hence,   $u_\x[i]$ is 0  with probability $\pzero=1-\pone$ and $N$ with probability $\pone$. The vector containing the number of molecules transmitted from this TBN at various slots is denoted by $\mathbf{u}_\x=(\cdots,u_\x[-1],u_\x[0],u_\x[1],\cdots)$. To include the independence and randomness of information bits, the vector $u_\x$ is  assigned  to  this TBN as its mark. Hence, the interfering TBNs can be modeled using a marked PPP $\phim=\{(\x,\mathbf{u}_\x):\x\in\tbnproc\}$. Here, the mark $u_\x$ is independent of marks of other transmitters. \color{black}The transmitter and receiver are assumed to be synchronized in time. Without loss of generality, we consider the slot 0. 
Hence, $u_\x[0], u_\x[-1],$ $u_\x[-2]...u_\x[-L+1]$ would denote the number of IMs emitted by the tagged TBN  corresponding to the current bit $\sx[0]$, previous bit $\sx[-1]$ and up to the $(L-1)\ths$ previous bit $\sx[-L+1]$ respectively. We consider OOK modulation, therefore $u_\x[i]=N$ when $\sx[i]=1$ and $u_\x[i]=0$ otherwise for any $i\ths$ slot.


\subsection{Propagation model} The propagation mechanism considered in this work is free diffusion. The IMs emitted from the point source to the fluid medium propagate to the receiver via 3D Brownian motion.  
The concentration of IMs is assumed to be small enough so that the collision between them can be neglected. This assumption allows us to consider the propagation of these molecules to be independent of each other. The diffusion coefficient can be considered to be constant  by assuming temperature and viscosity in the propagation environment as homogeneous and constant along with the above assumption \cite{channel}.

\subsection{Channel and receiving model}
Let us consider a TBN at location $\xt$ transmitting IMs. Let $\cir{t}{\norm{\xt}}$ denote the fraction of its IMs  reaching the RBN within time $t$ since the transmission. Hence, the fraction of IMs reaching the RBN in $l\ths$ slot  (\ie in the time window $[l\ts ,(l+1)\ts ]$) since the transmission due to impulsive emission of IMs (at the begining of time slot) at the TBN, is given as
\begin{align}
\cirslot{\norm{\xt}}{l}=\cir{(l+1)\ts}{\norm{\xt}}-\cir{l\ts}{\norm{\xt}}\label{s341}.
\end{align}
$\cirslot{\norm{\xt}}{l}$ is  known as channel impulse response (CIR), which is defined as the probability of observation of one molecule at time
$t$ at the receiver when the transmitter is stimulated in an impulsive manner at time $t_0 = 0$ \cite{channel}.
Consider the arrival of IMs  at the RBN at  $l\ths$ time slot since the transmission  as Bernoulli trials with probability of success $\cirslot{\norm{\xt}}{l}$. If the number of transmitted IMs is $N$, then  the number of molecules $y_{\xt}$ observed at the RBN at the $l\ths$ slot since transmission follows a Binomial distribution with parameter $(N,\cirslot{\norm{\xt}}{l})$. Assuming $N$ is large  and success probability $\cirslot{\norm{\xt}}{l}$ is small, we can approximate Binomial distribution with Poisson distribution for mathematical tractability \cite{channel}. Hence $y_{\xt}\sim\mathcal{P}(N\cirslot{\norm{\xt}}{l})$.
The total number of desired IMs (\ie IMs emitted from the tagged TBN) reaching RBN at current slot (\ie slot 0) is given as
\begin{align}
y_\mathrm{S}\sim\mathcal{P}\left( \cirslot{\norm{\xd}}{0}u_{\xd}[0]\right)\label{eq:yS}.
\end{align} 

All IMs emitted by the tagged TBN at the current time slot may not reach the receiver at the current time slot itself. The remaining molecules wander in the environment and may arrive at RBN in the later slots. 
The total number of IMs (that were emitted at slot $[-l]$ from the tagged TBN) reaching the RBN at slot 0 is distributed as 
$
\mathcal{P}\left(
\cirslot{\norm{\xd}}{l} u_{\xd}[-l]
\right)
$. 
Since the sum of Poisson random variables is Poisson itself, the observed number of molecules from all the previous slots is given as
\begin{align}
y_\mathrm{I}\sim\mathcal{P}\left( \sum_{l=1}^\infty\cirslot{\norm{\xd}}{l}u_{\xd}[-l]\right)\label{eq:yI}.
\end{align} 
Note that a fraction of molecules may never reach the RBN and only a fraction of the transmitted molecules reach the receptors of the RBN.
Since receivers cannot identify whether the molecules are from current or previous time slots, it results in ISI.

Similarly, the molecules transmitted from the interfering TBNs at the current and previous slots also mix with the molecules emitted by the tagged TBN to cause CCI. 
Similar to the case of $y_\mathrm{S}$ and $y_\mathrm{I}$, the total number of these interfering molecules, given $\tbnproc$, is given as
\begin{align}
y_\mathrm{C}&\sim \mathcal{P}\left( \sum_{\x\in \tbnproc}\sum_{l=0}^{\infty}\cirslot{\norm{\x}}{l}u_{\x}[-l]\right)\label{eq:yM}.
\end{align}
Thus the received signal  (\ie total number $y$ of IMs received) at the RBN at any time instant is the sum of the desired signal, ISI and CCI. 

\subsection{Decoding at RBN}
Let the detector used at the receiver be a threshold detector. At the end of the time slot, the RBN counts the number of molecules absorbed ($y$) in that time slot, and for demodulation, it is compared with a predefined threshold $\eta$. If $y<\eta$, then the transmitted bit $s_{\xd}[0]$ from the desired transmitter is estimated as $\hat{s}_{\xd}[0]=0$, otherwise $\hat{s}_{\xd}[0]=1$. An error would occur when the transmitted bit $s_{\xd}[0]=0$ is decoded as $\hat{s}_{\xd}[0]=1$ and vice versa. Therefore, the total probability of bit error ($\probe$) is given by
\begin{align}
\probe= &\pzero \probeb{0}+\pone \probeb{1}\label{eq:berdef}
\end{align}
where $\probeb{0}$ and $\probeb{1}$ are the probability of incorrect decoding for bit 0 and 1, formally defined as
\begin{align}
\probeb{0}&=
\prob{
	\hat{s}_{\xd}[0]=1\mid s_{\xd}[0]=0
}
\\
\probeb{1}&=
\prob{
	\hat{s}_{\xd}[0]=0\mid s_{\xd}[0]=1
}.
\end{align}

\subsection{Modeling molecular degradation} Employing biodegradable IMs can improve the performance of the molecular communication system. The time duration required for the molecular concentration to reduce to the half of its initial concentration is termed as half-life  ($\Lambda_{1/2}$) which varies for different types of molecules\cite{deg1}. 
Incorporating adequate amount of molecular degradation in the design reduces the concentration of the delayed interfering molecules and thereby improves the performance of the molecular communication system.
We consider exponential degradation where
the probability that a molecule will not degrade in time  $t$ is equal to $\exp{\left(-\mu t\right)}$. Here, $\mu$ denotes the reaction rate constant of molecular degradation which is related to the half-time as $\mu=\ln(2)/\Lambda_{1/2}$. When reaction rate tends to zero ($\mu\rightarrow 0$, {\em i.e. } half-time is infinity $\Lambda_{1/2}\rightarrow \infty$), molecules will never undergo degradation.
We also assume that the molecule does not get involved in any other reactions. 

\subsection{Channel impulse response} 
Recall that the considered RBN is a spherical fully-absorbing receiver with radius $a$ and is located at the origin. Consider a point transmitter located at $r$ distance away from the center of the receiver. Then, the hitting rate of molecules  at the surface of the receiver (\ie total number of molecules hitting the receiver in unit time) at time $\tau$ is given as \cite{lecbio},
\begin{equation}
\kappa(\tau, r)=\frac{a}{r}\frac{r-a}{\sqrt{4\pi D \tau^3}}\exp\left( -\frac{(r-a)^2}{4D\tau}\right) ,\label{s31}
\end{equation}
where $D$ represents the diffusion coefficient, which depends on the properties of molecule used and the propagation medium.  Now, the fraction of non-degraded information molecule reaching the receiver within time $t$, is given by \cite{heren2015effect},

\begin{align}
\cir{t}{r}x=\int_{0}^{t}\kappa(\tau, r)\times \expS{-\mu \tau}\dd\tau
=& \frac{a}{2r}\left[\exp\left(-\sqrt{\frac{\mu}{D}}(r-a)
\right)\erfc{\frac{r-a}{\sqrt{4Dt}} -\sqrt{\mu t}}\right.\nonumber\\&
\left.
+\exp\left(\sqrt{\frac{\mu}{D}}(r-a)
\right)\erfc{\frac{r-a}{\sqrt{4Dt}} +\sqrt{\mu t} }\right]\label{s32}.
\end{align}
\begin{remark}
	The fraction of IMs eventually reaching the RBN is 
	\begin{align}
	f(\infty, r)=\frac{a}{r}\expS{-\sqrt{\frac{\mu}{D}}(r-a)}.\label{s33a}
	\end{align}
\end{remark}

\begin{remark}
	The fraction of IMs reaching the RBN within time $t$ when there is no degradation is \cite{yilmaz2014three}
	\begin{equation}
	f_0(t, r)=\lim\limits_{\mu \rightarrow 0}f(t, r)=\frac{a}{r}\erfc{ \frac{r-a}{\sqrt{4Dt}}}.\label{s33}
	\end{equation}
\end{remark}


\section{Observations at the RBN}

Total number of IMs received at the RBN at the current slot 0 (including all sources and previous slots) is given as
\begin{align*}
y=y_\mathrm{S}+y_\mathrm{I}+y_\mathrm{C}.
\end{align*}
Using \eqref{eq:yS},\eqref{eq:yI},\eqref{eq:yM} and noting that the sum of Poisson random variables is also a Poisson random variable, 

\begin{align}
y&\sim \mathcal{P}\left( \sum_{\xt\in \phitotal}\sum_{l=0}^{\infty}h_{\norm{\xt}}[l]u_{\xt}[-l]\right).\label{s45}
\end{align}
\par
Now, from \eqref{eq:yS}, the expected number of desired signal IMs reaching the RBN at the current time slot is given by,

\begin{align}
\Es &=\expect{y_{\mathrm{S}}}=
\pone N\cir{\ts}{\rd}.\label{eq:averageES}
\end{align}

Similarly, from \eqref{eq:yI}, the expected number of interfering molecules from the desired transmitter absorbed at the current time slot is given by

\begin{align}
\Ei =\expect{y_{\mathrm{I}}}
&=\pone N
\sum_{l=1}^{\infty}h_{\norm{\xd}}[l]= \pone N\left(\cir{\infty}{\rd}-
\cir{\ts}{\rd}\right)\label{eq:averageEI}
\end{align}

where the last step is obtained by substituting the value of $h_{\norm{\xd}}[l]$  from \eqref{s341}.

The expected number of molecules arriving at the current time slot from the interfering TBNs 
is given by (See Appendix \ref{app:A} for proof.)

	\begin{align}
	\Em &=\exps\left[y_{\mathrm{C}}\right]=4\pi\lambda \pone Na\left(\frac{D}{\mu}+a\sqrt{\frac{D}{\mu}}\right).\label{eq:averageEM}
	\end{align}

%
It can be seen that the expected CCI (number of  IMs that were emitted by the interfering TBNs and absorbed at the RBN) 
increases with  $\lambda , \pone, N , a$ and $D$, and decreases with $\mu$. Also in \eqref{eq:averageEM}, we can see that when $\mu\rightarrow 0$ (no molecular degradation), $\Em \rightarrow \infty$. That is, for a system with no molecular degradation, the expected CCI will tend to infinity at the steady state. 
\par
 In the system's transition state, when there has been only $K-1$ previous transmissions before the current slot, the expected number of interfering molecules absorbed at the RBN from the desired transmitter is
	\begin{align}
	\mathrm{E_I^{Transient}} &=\pone N
	\sum_{l=1}^{K-1}h_{\norm{\xd}}[l]= \pone N\left(\cir{K\ts}{\rd}-
	\cir{\ts}{\rd}\right)\label{en16}
	\end{align}
	The expected CCI, in case of no degradation for the system at the transient state, is (See Appendix \ref{app:B} for proof)
	\begin{align}
	\mathrm{E_C^{Transient}}=4\lambda \pi \pone Na\left(DK\ts +a\sqrt{\frac{4DK\ts }{\pi}}\right).\label{s49}
	\end{align}
Now from \eqref{eq:averageES}, \eqref{eq:averageEI} and \eqref{eq:averageEM}, the expected total number of IMs absorbed by the RBN at any time slot is 

\begin{align}
\Et &=\Es +\Ei +\Em =\pone N a \left( \frac{1}{r}\expS{-\sqrt{\frac{\mu}{D}}(r-a)}\right.\left. +4\pi\lambda \left(\frac{D}{\mu}+a\sqrt{\frac{D}{\mu}}\right)\right)\label{eq:averageET}.
\end{align}
\section{Probability of bit error}
In this section, we derive the probability of bit error as defined in \eqref{eq:berdef} for the considered molecular communication system. 
We will first consider systems where ISI is negligible and 
 then, extend the analysis to the systems with ISI. 
\subsection{System without ISI}
In this subsection, we will consider a system where ISI is negligible. 
Some examples of such system include cases where the symbol time $\ts $ is sufficiently large and/or molecular degradation rate $\mu$ is sufficient (see Fig. \ref{fig:f3}). 
For a system without ISI, the number of molecules received at the typical RBN is the sum of IMs corresponding to the current symbol of tagged TBN, and IMs from other interfering TBNs corresponding to the current slot. Therefore, conditioned on $\phim$, the number of absorbed molecules observed at the receiver is
\begin{align}
y\sim\mathcal{P}\left(\overbrace{h_{\norm{\xd}}[0]u_{\xd}[0]}^{\text{Desired}}+\overbrace{\sum_{\x\in \phim}h_{\norm{\x}}[0]u_{\x}[ 0]}^{\text{CCI}}\right).
\end{align}

We will first fix the distance $\rd$ between the tagged TBN and the RBN and derive $\probe$. We will, then, derive $\probe$ for an arbitrary distribution of $\rd$.  

\noindent\textbf{Case I: When $\rd$ is constant:} \\
The probability of bit error for the case when $\rd$ is constant is given in Theorem \ref{thm:fixedrdnoisi}.
\begin{theorem}\label{thm:fixedrdnoisi}
For a system with no ISI, the probability of bit error  is given by \eqref{eq:berdef} with the probability of incorrect decoding of bit 0 and 1  given as
\begin{align}
\probeb{0}&=1-\expU{-\alpha_0(\lambda)} 
\left[ 			
1+\sum_{n=1}^{\eta-1}\frac{1}{n!}\mathfrak{B}_n (\pmb{\alpha}(\lambda))
\right],
\label{eq12}\\
\probeb{1}&=
\expU{-\alpha_0(\lambda)} \expS{-  N\cir{\ts}{\rd}
}\times \left[1+\sum_{n=1}^{\eta-1}\frac{1}{n!}\mathfrak{B}_n (\pmb{\beta}(\rd,\lambda))\right],\label{eq13}
\end{align}
where
\begin{align*}
\alpha_0(\lambda)=4\pi \lambda \pone \int_{a}^{\infty}
	\left[
	1-\expU{ - N\cir{\ts}{z}}
	\right]
	z^2\dd z,
\end{align*} $\pmb{\alpha}(\lambda)=[\alpha_{1}(\lambda),\alpha_{2}(\lambda),...,\alpha_{\eta-1}(\lambda)]$ 
and $\pmb{\beta}(\rd,\lambda)=[\alpha_{1}(\lambda)+N\cir{\ts}{\rd},\alpha_{2}(\lambda),...,\alpha_{\eta-1}(\lambda)]$ with
\begin{align}
&\alpha_i(\lambda)= 4\pi \lambda \pone \int_{a}^{\infty}
\expU{
	-N\cir{\ts}{z}
}
 {\left(N\cir{\ts}{z}\right)}^i 
 z^2\dd z.
\end{align}
Here, $\mathfrak{B}_n(.)$ denotes the $n$th complete exponential Bell's polynomial \cite{comtet2012advanced} given as
\begin{align}
&\mathfrak{B}_{n} (\pmb{\alpha}(\lambda))=\sum_{w=1}^n\sum\frac{n!}{j_1!j_2!...j_{n-w+1}!} \prod_{v=1}^{n-w+1}\left(\frac{\alpha_v(\lambda)}{v!}\right)^{j_v}.\label{pbp}
\end{align}
where the second sum is taken over all non-negative integers $j_1,j_2,...,j_{n-w+1}$ such that $j_1+j_2+...+j_{n-w+1}=w$ and $1j_1+2j_2+...+(n-w+1)j_{n-w+1}=n$.
\end{theorem}
\begin{IEEEproof}
	See Appendix \ref{app:C}.
\end{IEEEproof}
\begin{coro}
	When detection threshold $\eta=1$ and the 
	 bit 1 and 0 are equi-probable, 
	$\probe$ is given as 
\begin{align}
	\probe&=\frac12\left[
1-\expS{-2\pi \lambda\int_{a}^{\infty}
	\left[
	1-\expU{- N\cir{\ts}{z}}
	\right]
	z^2\dd z}\right.\left.\times\left(1-\expU{- N\cir{\ts}{\rd}}\right)\right].
\end{align}
\end{coro}
\begin{remark}
	Note that when $\eta=0$, the receiver will always decode $\cb=1$. Hence, $\probeb{0}=1$, $\probeb{1}=0$ and $\probe=\pzero$. As detection threshold $\eta$ increases, $\probeb{0}$ monotonically decreases, while $\probeb{1}$ monotonically increases. As $\eta\rightarrow \infty$,  $\probeb{0}=0$, $\probeb{1}=1$ and $\probe=\pone$. Since $\probe=\pone\probeb{1}+\pzero \probeb{0}$, there would be a trade-off   resulting in the existence of an optimal $\eta=\opt{\eta}$ for which $\probe$ is minimum. 
\end{remark}
\begin{remark}
	Note that $\probeb{0}$ does not depend on the value of $\rd$. Owing to the decreasing nature of $\cir{\ts}{\rd}$ with $\rd$ for large $\ts$ (see \eqref{s33a}), it can be shown via coupling argument \cite{lindvall2002lectures} that $\probeb{1}$ increases with $\rd$ for a given $\eta$. Hence, $\eta$ should be decreased when $\rd$ increases to maintain the same level of the probability of bit error.
\end{remark}
\noindent\textbf{Case-II: When $\rd$ is a random variable:}\\
Now, we consider the case when $\rd$ is a random variable with probability distribution function  $\dist{\rd}{r}$. This case is more realistic as the tagged TBN is not fixed and can move in the medium. Example  includes the uniform distribution where the transmitter distance is uniformly distributed between $b$ and $c$ such that $a<b<c$. Hence, 
$\dist{\rd}{r}={1}/{(c-b)}\1(b\leq r\leq c)$. 
The probability of bit error for this case is given in Theorem \ref{thm:randomrdnoisi}.
\begin{theorem}\label{thm:randomrdnoisi}
	For a system with no ISI, the probability of bit error rate is given by \eqref{eq:berdef} with the probability of incorrect decoding for bit 0 and 1 given as 
	
		\begin{align}
	\probeb{0}&=1-\expU{-\alpha_0(\lambda)} 
	\left[ 			
	1+\sum_{n=1}^{\eta-1}\frac{1}{n!}\mathfrak{B}_n (\pmb{\alpha}(\lambda))
	\right],
	\label{eq20}\\
	\probeb{1}&=	
	\expU{-\alpha_0(\lambda)} 
	\int_0^\infty
	\expS{-  N\cir{\ts}{\rd}
	}\times \left[1+\sum_{n=1}^{\eta-1}\frac{1}{n!}\mathfrak{B}_n (\pmb{\beta}(\rd,\lambda))\right]
	\dist{R_\mathrm{d}}{\rd}\dd \rd\nonumber
	\\
	&
	=\expU{-\alpha_0(\lambda)} \left[\int_{0}^{\infty}
	\expU{- N\cir{\ts}{\rd}}
	\dist{}{\rd}\dd\rd+\right.
\sum_{n=1}^{\eta-1}\frac{1}{n!}
	\sum_{i=0}^{n}\binom{n}{i}\mathfrak{B}_{n-i} (\pmb{\alpha}(\lambda))\nonumber\\
	&\times\left.
	\int_{0}^{\infty}
	\expU{- N\cir{\ts}{\rd}}
	{\left(N\cir{\ts}{\rd}\right)}^i
	\dist{}{\rd}\dd\rd\right],\label{eq21}
	\end{align}
where  $\alpha_0(\lambda),\ \pmb{\alpha}(\lambda)$ and $\pmb{\beta}(\rd,\lambda)$ are the same as in Theorem \ref{thm:fixedrdnoisi}.
\end{theorem}
\begin{pf}
	See Appendix \ref{app:D}.
\end{pf}

\subsection{System with ISI}
We now consider systems with ISI. For simplicity, we will assume that ISI is limited to previous $L-1$ slots and the interference due to transmission in slots prior to $L$ slots is negligible.
We will first consider that the tagged  transmitter is at a fixed location $\xd$ and hence, $\rd$ is constant. The bit error rate for this system is given in Theorem \ref{thm:fixedrdisi}.

\begin{theorem}\label{thm:fixedrdisi}
	For the system with ISI from  $L-1$ previous slots, the probability of bit error is given by \eqref{eq:berdef} with the probability of incorrect decoding for bit 0 and 1 given as 
	\begin{align}
&\probeb{0}=1-\expS{
	-4\pi\lambda\int_{a}^{\infty}
	\left(
	1-\prod_{l=0}^{L-1}
	\efunc{l}{0}{z}
	\right)\ z^2\dd z	
}
\nonumber\\
&\times
\sum_{n=0}^{\eta-1}
\sum_{\sum_{i=1}^L m_i=n}
\frac{\mathfrak{B}_{ m_L} (\pmb{\epsilon}(\lambda))}
{ m_1! m_2!... m_L!}
\left(
\prod_{l=1}^{L-1}
\efunc{l}{m_l}{\rd}
\right),\label{eq24}\\
&\probeb{1}=\expU{- \cirslot{\rd}{0}N}
\expS{-4\pi\lambda\int_{a}^{\infty}
	\left(1-\prod_{l=0}^{L-1}
	\efunc{l}{0}{z}
	\right)\ z^2\mathrm{d}z}
\nonumber	\\	&\times
\sum_{n=0}^{\eta-1}
\sum_{ \sum_{i\in[1:L]} m_i=n}
\frac{\mathfrak{B}_{ m_L} (\pmb{\theta}(\rd,\lambda))}{ m_1! m_2!... m_L!}
\left(
\prod_{l=1}^{L-1}
\efunc{l}{m_l}{\rd}
\right).\label{eq25}
\end{align}

Here, $\pmb{\epsilon}(\lambda)=[\epsilon_{1}(\lambda),\epsilon_{2}(\lambda),\cdots \epsilon_{\eta-1}(\lambda)]$ and  $\pmb{\theta}(\rd,\lambda)=[\epsilon_{1}(\lambda) +\cirslot{\rd}{0}N,\epsilon_{2}(\lambda),\cdots \epsilon_{\eta-1}(\lambda)]$ with

	\begin{align}
	&\epsilon_{i}(\lambda)=4\pi \lambda \int_{a}^{\infty}\sum_{\sum_{0}^{L-1}q_j=i}\frac{i!}{q_0!q_1!...q_{L-1}!}\prod_{l=0}^{L-1}
	e_{l,q_l}(z) z^2\dd z,
	\end{align}
where $e_{l,k}(z)$ 
 is given by,
	\begin{align}
	e_{l,k}(z)=
	\pzero \1(k=0)+\pone {(\cirslot{z}{l}N)}^{k} \expS{-\cirslot{z}{l}N}.
	\end{align}
\end{theorem}
	\begin{pf}
		See Appendix \ref{app:E}.
	\end{pf}

Note that in \eqref{eq24} and \eqref{eq25}, the exponential term and the terms consisting of Bell polynomials, are due to the CCI and the $\efunc{l}{\cdot}{\rd}$ terms denote the interference due to $l\ths$ previous slot from the tagged TBN. 

We can derive the probability of bit error for a system with ISI and random $\rd$ 
in the same way  with an extra integral over the probability density function of $\rd$. The result is omitted here due to space limitation.

\section{Numerical Results}

In this section, we validate the derived analytical expressions 
by comparing them  with the corresponding Monte Carlo based simulations and present design insights about the molecular communication system with the help of numerical results. In all figures, the curves corresponding to the derived analytical expressions are represented by solid lines, and the simulation results are represented by markers.

For simulation, the interfering TBNs are generated as PPP outside the receiver volume up to a distance of $150\mu m$ from the center of the receiver. The interfering transmitters are distributed as PPP in the environment in each realization and simulation is done for $10^4$ realizations. The interfering TBN densities chosen for simulations are $10^{-5}$ and $10^{-4}$ TBNs per $\mu m^3$. This corresponds to 141 and 1414 interfering transmitters respectively. For all numerical results in this paper (unless stated otherwise), the diffusion coefficient is fixed as $D= 74.9 \ \mu m^2/s$, fully-absorbing spherical receiver radius is fixed as $a= 4\mu m$, the number of molecules emitted for bit-$1$ is $N=100$ molecules and the duration of the time slot is set as $0.5$ $s$. The above value of diffusion coefficient corresponds to the diffusion of human insulin-like molecule in blood like fluid at a temperature of 310$^\mathrm{o}$ K (body temperature).



\resultheading{Impact of the symbol time $(\ts)$ on the mean number of  IMs  received at the RBN $(\Es,\Ei,\Em$ and $\Et)$} 
\begin{figure}
	\centering
	\includegraphics[width=0.6\linewidth]{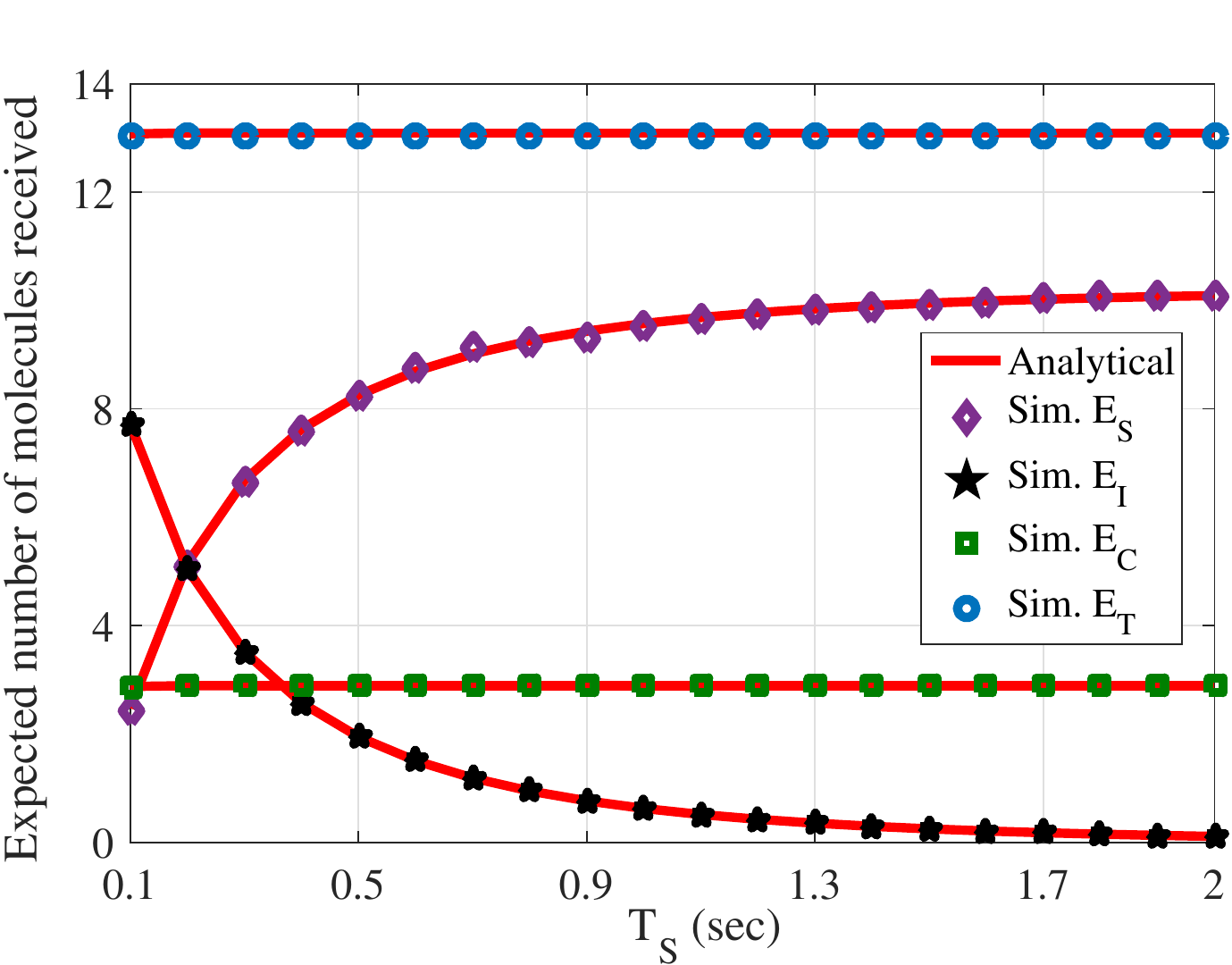}
	\caption{The expected number of desired ($\Es $), ISI ($\Ei $), CCI ($\Em $) and total ($\Et =\Es +\Ei +\Em $) IMs absorbed at the RBN for a system with  molecular degradation  versus symbol time ($\ts$). Here $\mu=1\text{s}^{-1},\  \rd=10\mu m \ \text{and} \ \lambda = 1\times 10^{-5} \text{TBNs}/ \mu m^3$. }
	\label{fig:f3}
\end{figure}
\begin{figure}
	\centering
	\includegraphics[width=0.6\linewidth]{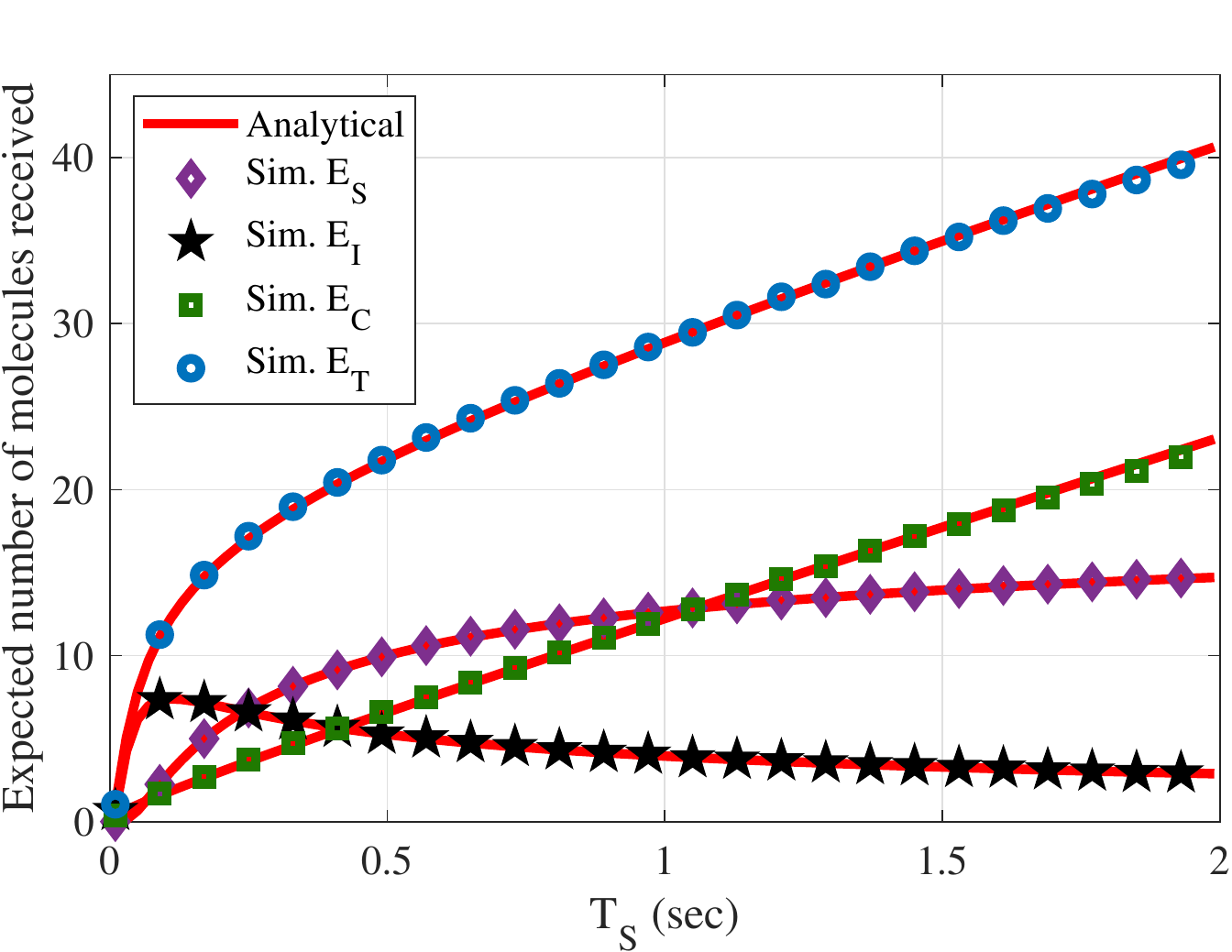}
	\caption{The expected number of desired, ISI, CCI, and total IMs absorbed at the RBN for a system without molecular degradation  versus symbol time ($\ts$) when the system is in its transient state (start time is 0). Here $\mu=0\text{s}^{-1},\  \rd=10\mu m \ \text{and} \ \lambda = 1\times 10^{-5} \text{TBNs}/ \mu m^3$.}\label{f1_tran}
\end{figure}
Fig. \ref{fig:f3} shows the variation of $\Et$ for a system with degradation rate $\mu=1\text{s}^{-1}$ with respect to $\ts$. Fig. \ref{fig:f3} is plotted using the steady state equations of $\Es,\Ei,\Em$ and $\Et$ derived in \eqref{eq:averageES}, \eqref{eq:averageEI}, \eqref{eq:averageEM} and \eqref{eq:averageET}.  The tagged TBN is at a fixed distance $\rd=10\mu m$ and interfering TBN density is set as $\lambda=1\times10^{-5}$ TBNs per $\mu m^3$. As seen in Fig. \ref{fig:f3}, $\Et$ is constant at steady state, which is also shown  in \eqref{eq:averageET}. Also the summation of the expected number of desired current slot IMs $(\Es)$ and previous slots IMs  $(\Ei)$ is also constant. As symbol time increases, the total number of IMs reaching RBN in their transmission slot only increases, hence $\Es$ increases. Therefore, the expected ISI ($\Ei$) decreases with symbol time $\ts$. Hence, it is important to select appropriate $\ts$ to reduce ISI and to improve the system's performance. It can also be observed that most of the ISI is  contained in some finite time duration (or finite number of time-slots which may depend on system's parameters) and hence, ISI from time after this duration can be neglected.\par
Fig. \ref{f1_tran} shows the variation of the expected total IMs absorbed by the RBN at its transient state for a system without molecular degradation. Fig. \ref{f1_tran}  is plotted using \eqref{eq:averageES}, \eqref{en16}, \eqref{s49} and $ \mathrm{E_T^{Transient}} =\mathrm{E_S}+\mathrm{E_I^{Transient}}+\mathrm{E_C^{Transient}}$. The expected total number of absorbed molecules increases with $ \ts $, and at the steady-state, the expected total number of absorbed molecules approaches infinite due to the flooding of molecules from the TBNs (can also be verified by substituting $ \mu=0 $s$^ {-1} $ in \eqref{eq:averageEM}).


\resultheading{Variation of $\Et$ with  the distance between the tagged TBN  and the RBN $(\rd)$}
\begin{figure}
	\centering
	\includegraphics[width=0.6\linewidth]{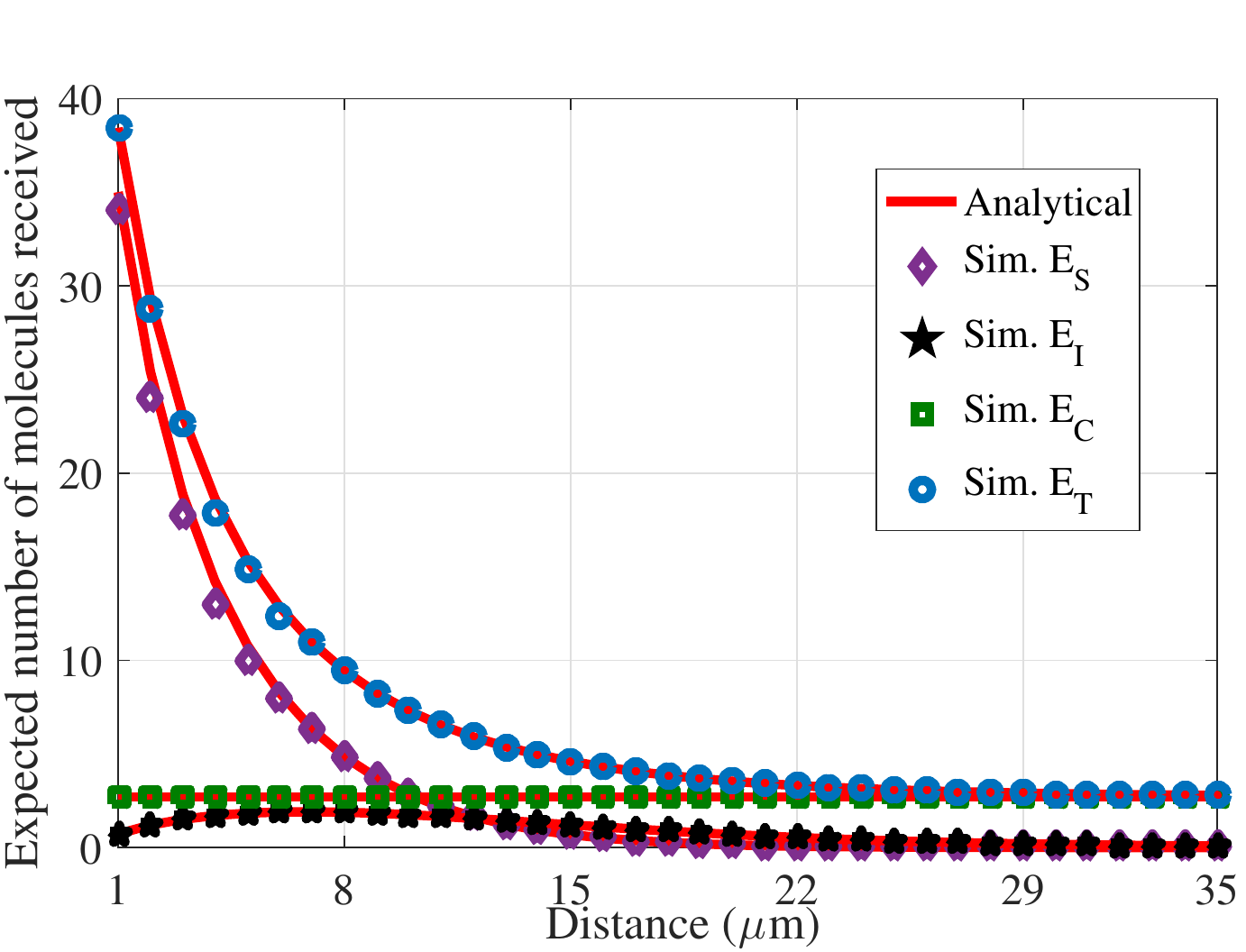}
	\caption{Variation of the expected number of desired ($\Es $), ISI ($\Ei $), CCI ($\Em $) and total ($\Et $) molecules absorbed at the receiver for a system with molecular degradation versus the distance between tagged TBN and the RBN's surface $(\rd-a)$. Here, $\mu=1\text{s}^{-1} \ \text{and} \ \lambda = 1\times 10^{-5} \text{ TBNs}/ \mu m^3$.}
	\label{fig:f4}
\end{figure}
Fig. \ref{fig:f4} shows the variation of $\Et$ with the distance between the surface of the spherical receiver and the tagged TBN, \ie $\rd-a$.  Fig. \ref{fig:f4} is plotted using the steady state equations of $\Es,\Ei,\Em$ and $\Et$ derived in \eqref{eq:averageES}, \eqref{eq:averageEI}, \eqref{eq:averageEM} and \eqref{eq:averageET}. The interfering TBN density is set as $\lambda= 1\times10^{-5}$ TBNs$/\mu m^3$ and the degradation rate constant $\mu=1\text{s}^{-1}$. The observation of the expected CCI IMs ($\Em$) at the receiver is independent of $\rd$  as seen in \eqref{eq:averageEM}. Hence, the tagged TBN's location  affects the observation of only desired IMs ($\Es$) and ISI ($\Ei$). $\Es$ reduces with the increase in $\rd$, however, $\Ei$ shows a non-monotonic behavior with  $\rd$. 
Owing the  combined behavior of $\Es,\Ei$ and $\Em$ with $\rd$,  $\Et$  reduces as the tagged TBN move away from the RBN. 
The result that $\Et$ varies with $\rd$,  indicates that decoding threshold $\eta$ should be chosen according to  $\rd$. 
At higher value of $\rd$, $\Es \  \text{and} \ \Ei$ reduces to zero and $\Et$ is only due to $\Em$ which will result in very high probability of bit error and a significant loss of information. We next discuss the impact of the threshold and the importance of selecting an appropriate threshold, for both cases when $\rd$ is fixed and when it is a random variable.  Recall that the decoding threshold  has an impact on the system's performance since the received bit is decoded as either 0 or 1 based on the threshold value.

\resultheading{Impact of decoding threshold $(\eta)$ on the probability of bit error $(\probe)$  when the location of tagged TBN is fixed}
\begin{figure}
	\centering
	\includegraphics[width=.6\linewidth]{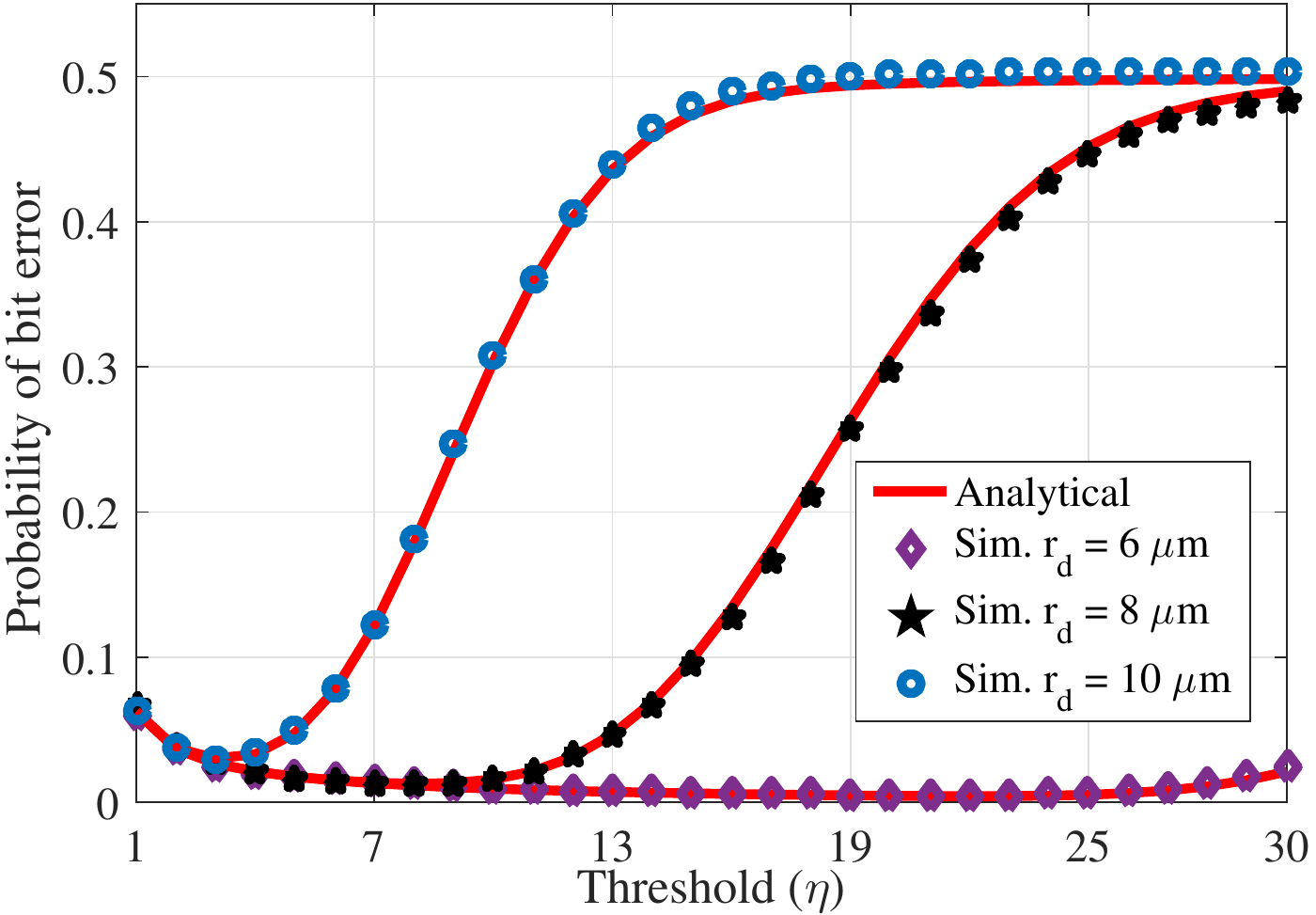}
	\caption{Probability of bit error ($\probe$) versus threshold ($\eta$) when $\rd$ is constant. When $\rd$ increases, $\probe$ increases while the optimal threshold $\opt{\eta}$ decreases. Here $\lambda = 1\times 10^{-5} \text{ TBN}/ \mu m^3 \ \text{and} \ \mu $ is set as $5\text{s}^{-1}$ to ensure ISI is negligible.}
	\label{fig:f5}
\end{figure}
We now compute $\probe$ for  the system with no or negligible ISI when the tagged TBN is fixed. 
$\mu$ is chosen to be $5\text{s}^{-1}$ to ensure that ISI is negligible. Fig. \ref{fig:f5} shows the variation of $\probe$ with the threshold $\eta$.
Analytical results (as derived in Theorem \ref{thm:fixedrdnoisi}) are compared with corresponding simulations for various values of $\rd$. It can be observed that as the threshold  is increased, $\probe$ first reduces, then reaches a minimum value and  increases after that. Therefore, there exists an optimum threshold $\opt{\eta}$ for which $\probe$ is minimum. When $\rd$ increases, $\opt{\eta}$ decreases 
owing to the reduction in the total number of IMs reaching the RBN. Due to the relative reduction in $\Es$, in comparison to $\Em$,  $\probe$ at $\opt{\eta}$ increases with $\rd$. 

\resultheading{Probability of bit error versus threshold for decoding when the desired transmitter distance is uniformly distributed}
\begin{figure}
	\centering
	\includegraphics[width=.6\linewidth]{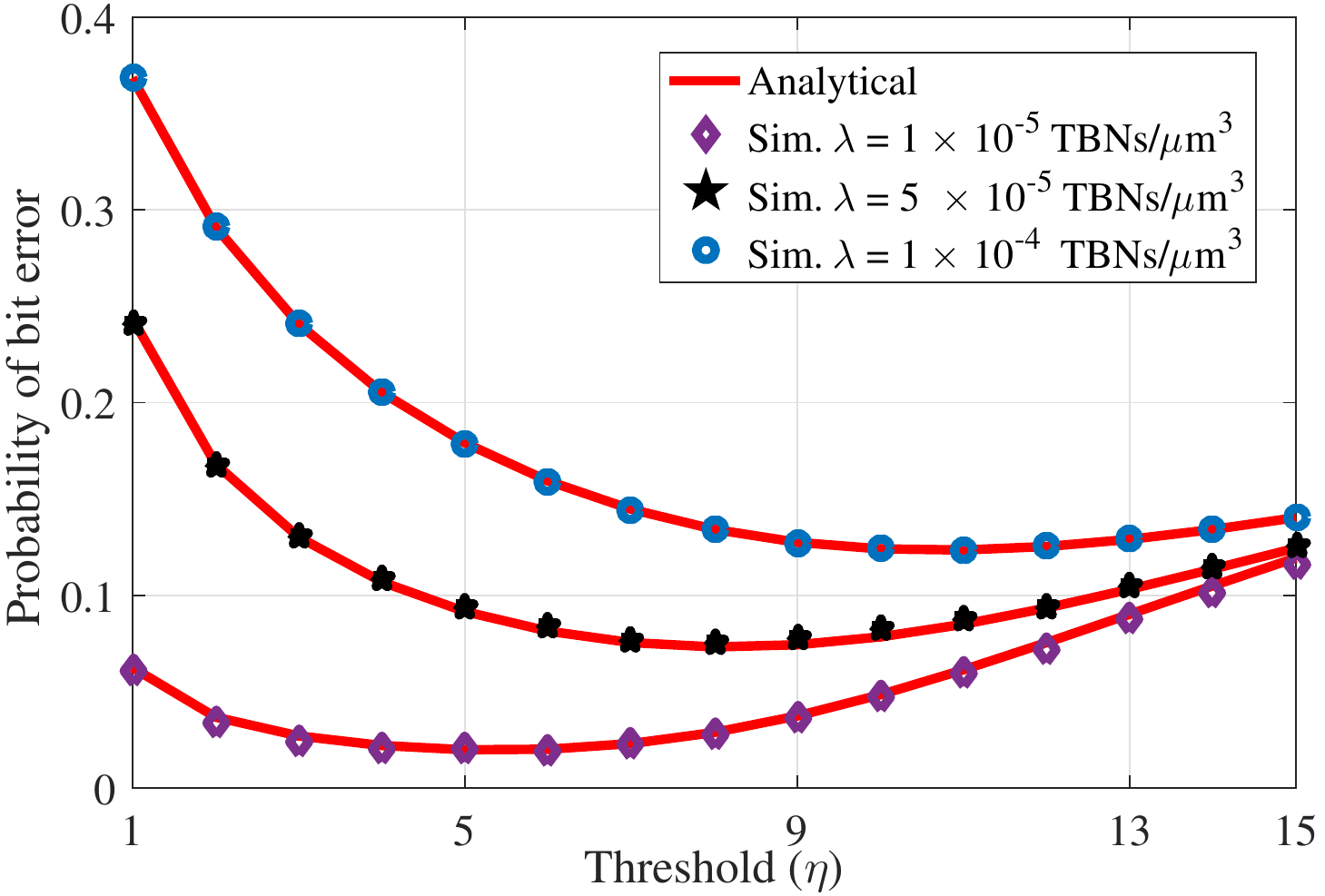}
	\caption{Probability of bit error versus the threshold $\eta$ with different interfering transmitter densities for a system with no ISI. Here, the desired transmitter is uniformly distributed between $b=4.1\mu m$ and $c=10\mu m$.}
	\label{fig:f6}
\end{figure}
Fig. \ref{fig:f6} shows the variation of probability of bit error with the threshold ($\eta$) when the distance of the desired transmitter is uniformly distributed, as derived in Theorem \ref{thm:randomrdnoisi}. The degradation rate  $\mu$ is set as $5\text{s}^{-1}$ to ensure that the ISI is negligible.  Here, $\rd$ is uniformly distributed between $b=4.1\mu m$ and $c=10\mu m$.  
Similar to Fig. \ref{fig:f5}, as $ \eta$ increases, $ \probe$ first decreases and achieves a minimum value and after that, $ \probe$ increases. Therefore, there exists an optimum threshold $\opt{\eta}$ for which $\probe$ is minimum. We have considered a single value of $ \eta$, irrespective of the value of $\rd$ for this result. We will discuss the impact of selecting $\eta$ according to $\rd$ in the next subsection. 
Fig. \ref{fig:f6} also shows the impact of the density ($\lambda$) of interfering TBNs. As $\Em$ increases with $\lambda$, $\probe$ also increases.

\resultheading{Gains from adaptive selection of decoding threshold}
\begin{figure}
	\centering
	\includegraphics[width=.6\linewidth]{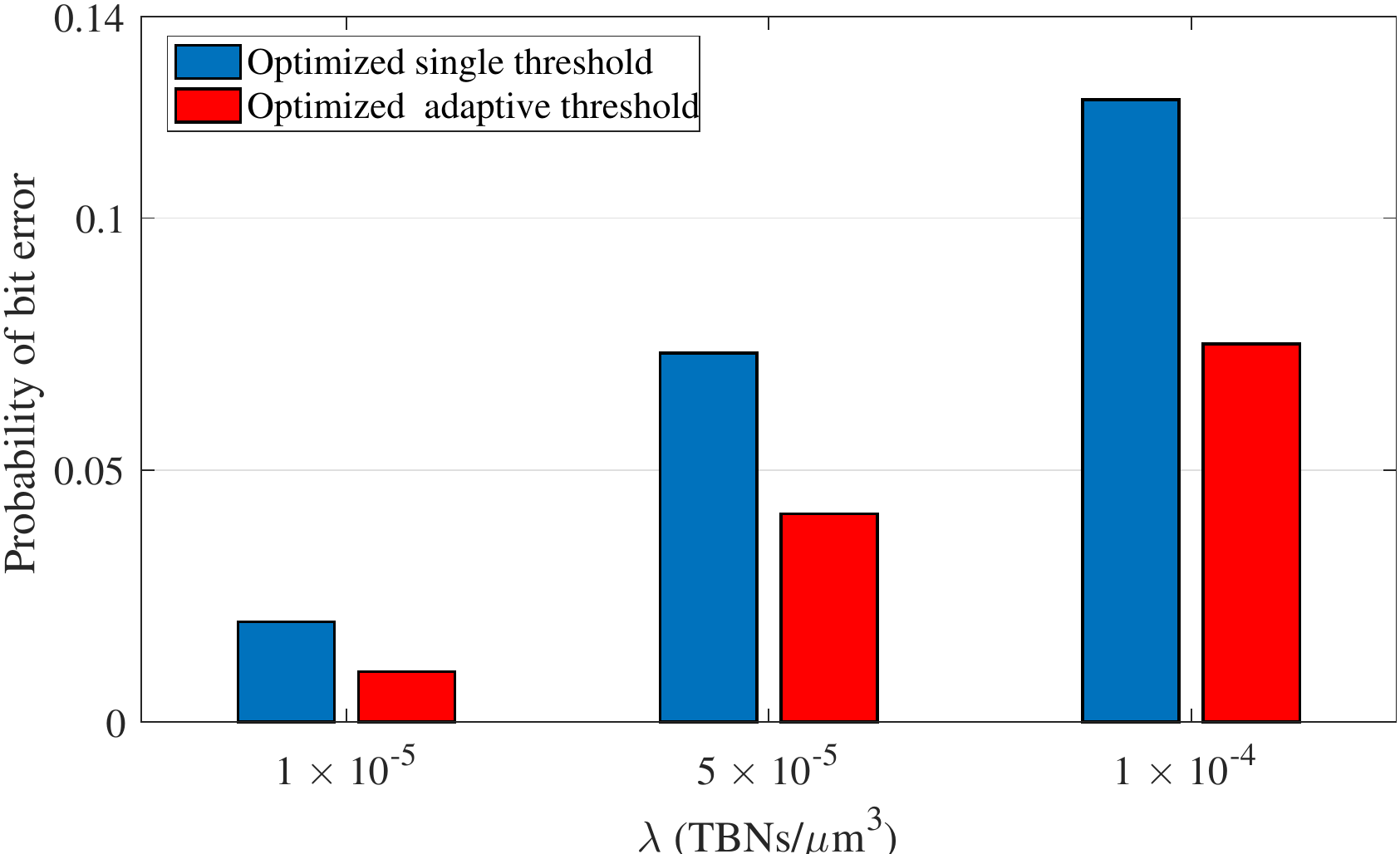}
	\caption{Comparison of the probability of bit error achievable with single threshold and adaptive threshold based decoding for different interfering transmitter densities for a system with no ISI. Here, $\mu=5\text{s}^{-1}$ and $\rd$ is uniformly distributed between $b=4.1\mu m$ and $c=10\mu m$.}
	\label{fig:f7}
\end{figure}
We now show the gains that can be achieved if the threshold can be adjusted according to the instantaneous channel. We consider the system where the distance between the tagged TBN and the RBN is 
uniformly distributed between $b=4.1\mu m$ and $c=10\mu m$. We compare two cases: the one where single optimal threshold $\eta$ is used and the case where threshold is chosen optimally according to the instantaneous value of $\rd$.   The optimal threshold is chosen from a database which was created by computing threshold values corresponding to minimum $\probe$ for a range of values for $\rd$ (with step size 0.1 $\mu$m) using \eqref{eq:berdef}, \eqref{eq12} and \eqref{eq13}. Assuming that an estimate for $\rd$ is available at the receiver, the decoder can use the threshold available in the database corresponding to this estimate. Fig. \ref{fig:f7} compares the bit error probability for the two cases for various values of density of interfering TBNs.
In the adaptive threshold-based decoding system, it  can be seen from Fig. \ref{fig:f7} that, there is a $40\%- 50\%$ reduction in the probability of bit error in comparison to the case when the single threshold-based decoding  is used.

\resultheading{Probability of bit error for the system with ISI}
\begin{figure}
	\centering
	\includegraphics[width=.6\linewidth]{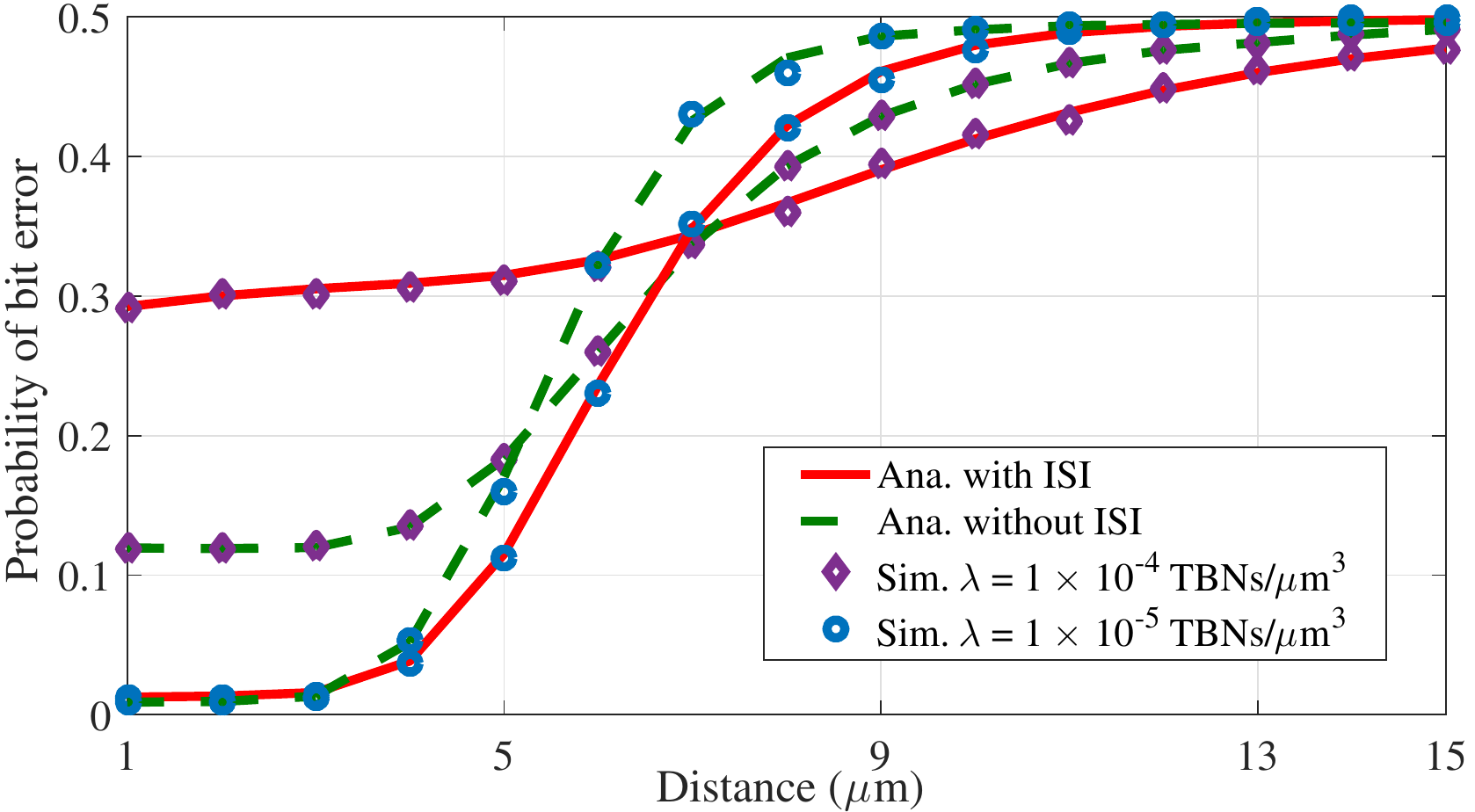}
	\caption{Probability of bit error versus the distance between the tagged TBN and the surface of the RBN \ie $(r_d-a)$ for two values of interfering TBN density in the presence of ISI. Here, $N=50$, $\mu=1\text{s}^{-1}$ and $\eta=10$. }
	\label{fig:f8}
\end{figure}
We now consider a system with ISI. We compute the probability of bit error for this system and show that accurately characterizing ISI is important to accurately compute the probability of bit error. Here, we consider $N=50$, $\mu=1\text{s}^{-1}$ and $L=5$. Value of $L$ is chosen such that ISI is contained in this time duration. The value of decoding threshold $\eta$ is taken as 10. 
Fig. \ref{fig:f8} shows the variation of probability of bit error with  ($r_d-a$),  computed using \eqref{eq:berdef}, \eqref{eq24} and \eqref{eq25} which are plotted using solid lines.
With the increase in $\rd$, $\probe$ increases due to reduction in $\Es$. 
Fig. \ref{fig:f8} also shows the probability of bit error for the  system when ISI is ignored (dotted curves).  These values are computed using Theorem \ref{thm:fixedrdnoisi}.  Fig. \ref{fig:f8} indicates that ignoring ISI will result in an  incorrect  probability of bit error.  Owing to the different levels of $\Em$ and $\Ei$ for ISI and no ISI case, the optimal threshold for these two cases may be different. Therefore, it is important to include ISI information while computing the optimal threshold to reduce the bit error probability. The ISI has a larger impact on the performance of systems with higher interfering TBN density.  It can be seen from presented results that, even-though molecular degradation improves system performance by reducing $\Ei$ and $\Em$ to a greater extent, for effective communication in a system with multiple interfering transmitters, the performance of such systems may not be good under some scenarios. Simple error-correcting codes, good decoding schemes, intelligent transmission methods etc. can be further used for improving the performance of such a system.

\resultheading{Importance of accurate modeling of the randomness of information bits}
\begin{figure}
	\centering
	\includegraphics[width=.6\linewidth]{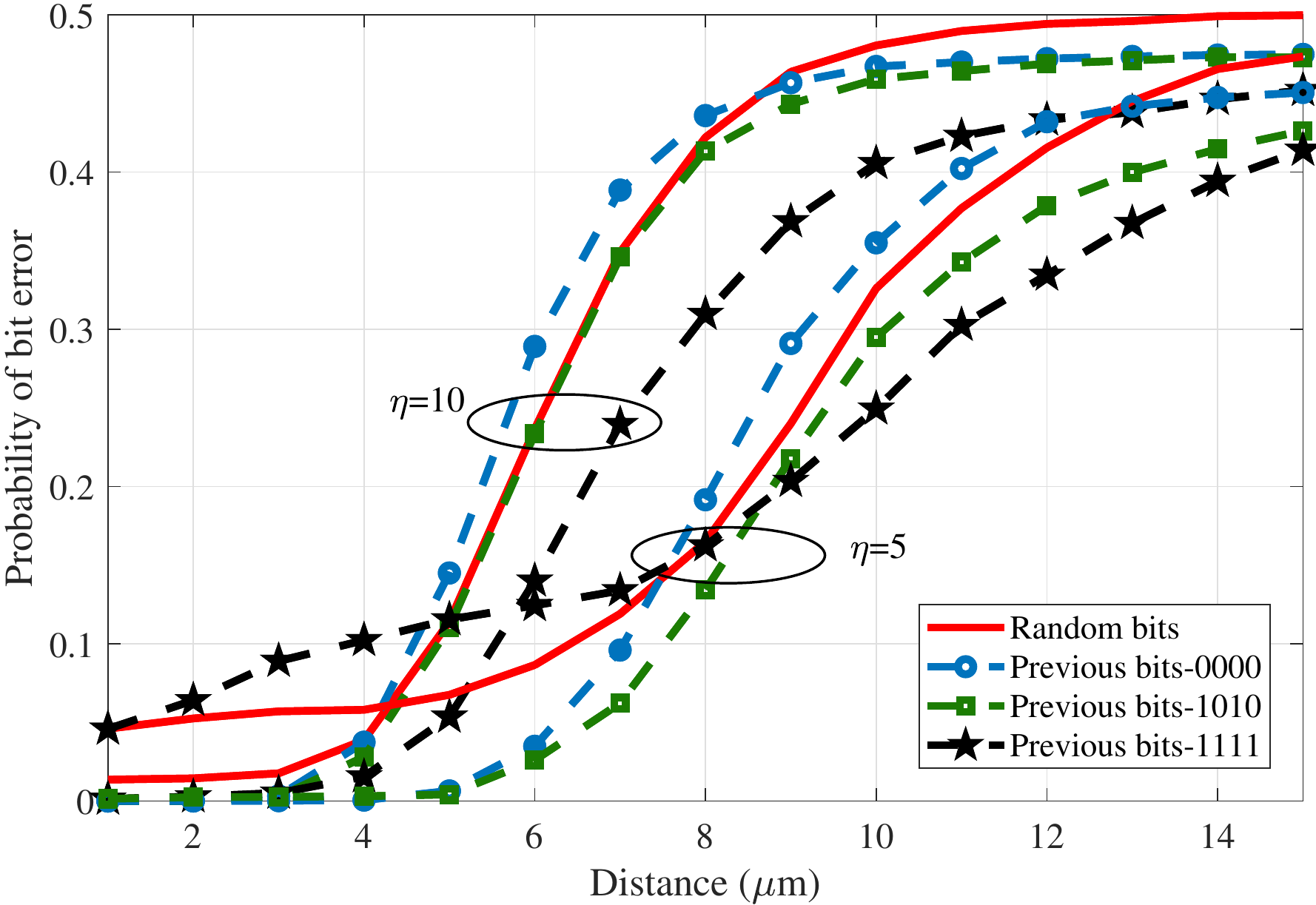}
	\caption{Comparison of the proposed model with model where  previous bits are fixed and all TBNs are transmitting the same bits. $\probe$  is plotted against  $(r_d-a)$.  Here $N=50$ molecules, $\lambda = 1\times 10^{-5} \text{TBN}/\mu m^3$, and $\mu=1\text{s}^{-1}$.}
	\label{fig:f9}
\end{figure}
\begin{figure}
	\centering
	\includegraphics[width=.6\linewidth]{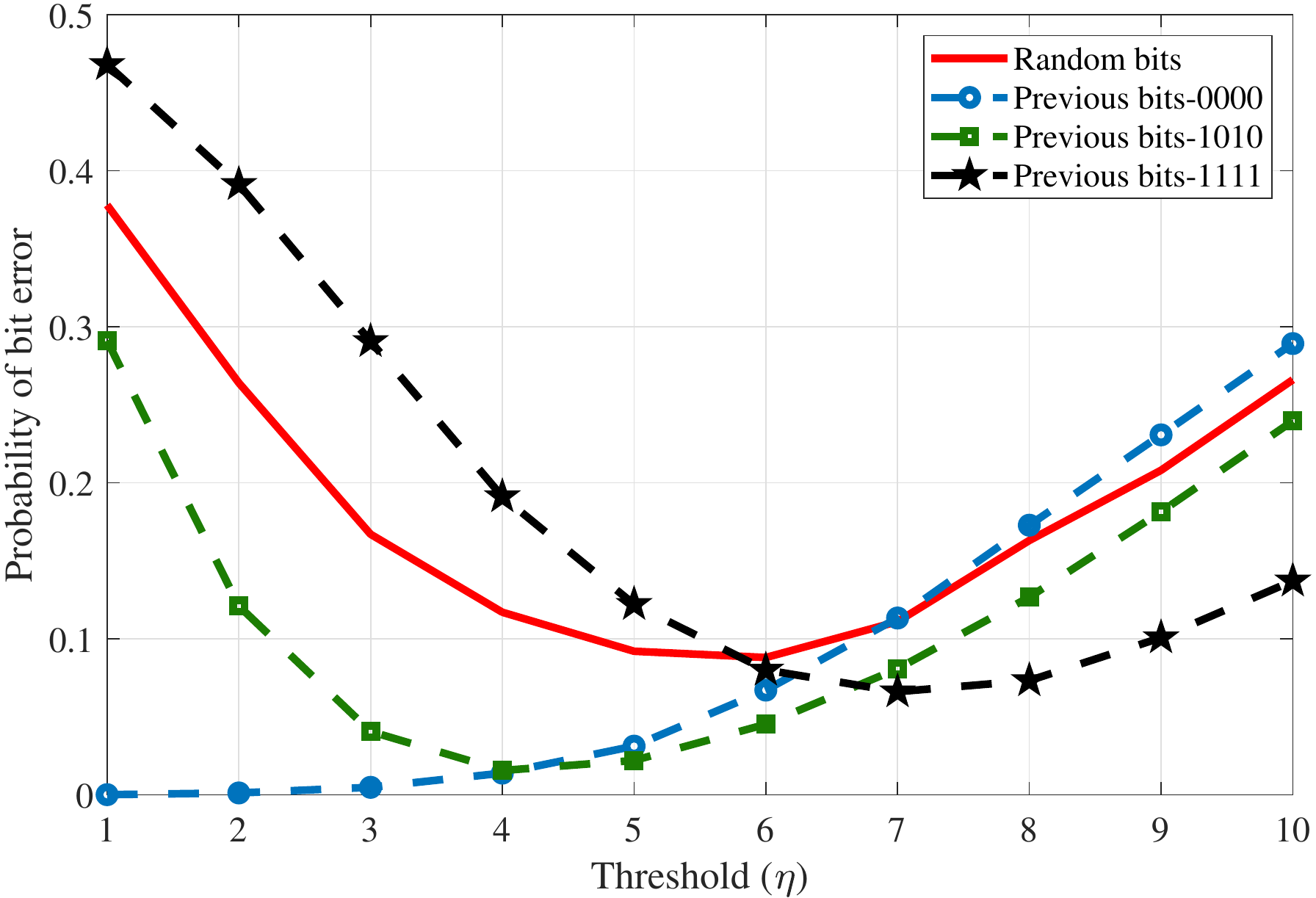}
	\caption{Variation of $\probe$ with the detection threshold $ \eta $ for the proposed model and  models where previous bits are conditioned and all TBNs are transmitting the same bits.  Here $N=50$ molecules, $\lambda = 1\times 10^{-5} \text{TBN}/\mu m^3$, $\mu=1\text{s}^{-1}$ and $\rd=10\mu$m.}
	\label{fig:f10}
\end{figure}
We now discuss why it is important to include the randomness and independence of information bits in the analysis. Recall that the past works have  derived the probability of bit error by considering  a conditioning  on  current and previous bits of tagged and interfering TBNs and by assuming the bits transmitted by all interfering TBNs in a slot is the same. In this paper, we have derived the  probability of bit error which system would see on the average when the transmit bits of a TBN are randomly  generated independent of other TBNs.  In Fig. \ref{fig:f9}, our proposed system is compared with corresponding simulation results of a system which is conditioned on the previous symbols and all the transmitters sending the same data. Fig. \ref{fig:f9} shows the variation of the probability of bit error with respect to the distance from the desired transmitter to the surface of the receiver. Here, $\lambda=1\times 10^{-5}$ TBN$/\mu m^3$,  $\eta = 5,\ 10$ and $\mu=1\text{s}^{-1}$.
 Similarly, Fig. \ref{fig:f10} shows the comparison of the proposed model with models where previous bits are fixed, and all TBNs are transmitting the same bits for different detection threshold $ \eta $. The chosen threshold determines which scenario is better out of the four scenarios considered. 
The  curves corresponding to the proposed work (red solid curve) is plotted using \eqref{eq:berdef}, \eqref{eq24} and \eqref{eq25}.
 Since the BER performance depends on the detection threshold and the optimal threshold is different for different cases, it is important to include the independence and randomness of the information bits in the analysis to compute the optimal threshold for a specific scenario. From the figures, it is evident that $\probe$ derived while assuming fixed previous symbols depends on the transmitted bit sequence and is  different from the $\probe$ of the real scenario (where all bits are random). Hence, considering the randomness and independence of information bits in the system model can provide us a gain in the accuracy of results.
\section{Conclusions}
In this paper, we have presented an analytical framework for a molecular communication system with multiple transmitters each having random transmit message independent of others. We have
derived the analytical expressions for expected number of molecules absorbed by the fully absorbing RBN that are contributing toward desired signal, ISI and CCI.
We have also derived the analytical expressions for the probability of bit error for the  systems by including the impact of interference from previous slots and other interfering TBNs. 
We have discussed the need of selecting threshold based on system parameters such as distance ($\rd$) between the RBN and the tagged TBN, the density of TBNs etc. We have also shown that  decoding using an adaptive threshold that is based on instantaneous channel condition (for example the instantaneous value of $\rd$) has the potential to reduce  the probability of bit error compared to single threshold-based decoding. We have shown the importance of accurately characterizing interference and including the randomness and independence of transmission bits in the analysis.

\appendices
\section{Derivation of the mean  CCI  at the RBN}\label{app:A}
 The expected number of interfering molecules observed at the RBN is given by,
\begin{align}
\Em &=\expect{y_{\mathrm{C}}}=\mathbb{E}_\tbnproc\left[\expect{y_{\mathrm{C}} \mid \tbnproc }\right].
\end{align}
Recall that \begin{align}
y_\mathrm{C}\mid \tbnproc&\sim \mathcal{P}\left( \sum_{\x\in \tbnproc}\sum_{l=0}^{\infty}\cirslot{\norm{\x}}{l}u_{\x}[-l]\right).
\end{align}
Hence,
\begin{align}
\Em &=\expect{y_{\mathrm{C}}}=
\mathbb{E}\left[\sum_{\x\in \tbnproc}\sum_{l=0}^{\infty}\cirslot{\norm{\x}}{l}u_{\x}[-l]]\right].\label{a1}	
\end{align}
Applying marked version of Campbell theorem \cite{haenggi2012stochastic} in \eqref{a1} gives,
\begin{align}
\Em &=4\pi\lambda \int_{a}^{\infty}\sum_{l=0}^{\infty}h_{z}[l]\mathbb{E}[u_{z}[-l]]z^2\mathrm{d}z\nonumber\\
&=4\pi\lambda \pone N
\int_{a}^{\infty}\sum_{l=0}^{\infty} h_{z}[l] z^2 \mathrm{d}z \label{eq:appA1}\\
&=4\pi\lambda  \pone N\int_{a}^{\infty} \cir{\infty}{z} z^2 \mathrm{d}z.\nonumber\\
&=4\pi\lambda \pone N\int_{a}^{\infty}a\exp\left(-\sqrt{\frac{\mu}{D}}(z-a)\right)z\dd z\nonumber\\
&=4\pi\lambda \pone Na\left(\frac{D}{\mu}+a\sqrt{\frac{D}{\mu}}\right)
	\label{a2}
\end{align}
\section{
}
\label{app:B}
For a system in transient state when only $K-1$ transmission has occurred in the past, $u_\x[l]=0$ for all $l\leq-K$. Hence, \eqref{eq:appA1} gives
\begin{align}
\Em &=4\pi\lambda \pone N
\int_{a}^{\infty}\sum_{l=0}^{K-1} h_{z}[l] z^2 \mathrm{d}z=4\pi\lambda  \pone N\int_{a}^{\infty} \cir{K\ts}{z} z^2 \dd z.
\end{align}
Now, using \eqref{s33}, we get

\begin{align}
\Em &=4\pi\lambda \pone  N\int_{a}^{\infty}a \ \erfc{ \frac{z-a}{\sqrt{4DK\ts }}}z \mathrm{d}z=4\lambda \pi \pone Na\left(DK\ts +a\sqrt{\frac{4DK\ts }{\pi}}\right).	\label{a3}
\end{align}
\section{Proof of Theorem-1}\label{app:C}
 Let us denote the current bit of the tagged TBN as $\cb$.  Hence, $u_{\xd}[0]=\cb N$. 
 
 Let $v(\norm{\x})$ be the expected number of IMs that were emitted by the transmitter located  at $\x$ and absorbed by the RBN. {\em i.e.},
\begin{equation}
v(\norm{\x})=\cirslot{\norm{\x}}{0} u_{\x}[ 0].
\end{equation}
Let $V$ be  the expected total number of received molecules conditioned on $\phim$ {\em i.e.},
\begin{align}
V(\rd,\cb,\phim)=\cb N\cirslot{\rd}{0}+\sum_{\x\in \phim} v(\norm{\x}).
\end{align}
Now, given $\phim$, the total number of IMs received is Poisson distributed \ie 
\begin{align}
y\mid \phim &\sim \mathcal{P}(V(\rd,\cb,\phim)).\label{eq:AppCymdist}
\end{align}
The probability of incorrect decoding for bit $\cb$ is given as
$
\probeb{\cb}=\prob{y\notin [\tbl{\cb}\ \tbh{\cb} ]},
$
where $\tbh{\cb}$ and $\tbl{\cb}$ are upper and lower limit of the decoding region of bit $\cb$. In particular, $\tbh{0}=\eta-1$, $\tbl{0}=0$, $\tbh{1}=\infty$, and $\tbl{1}=\eta$.  Now, 
the probability of incorrect decoding for bit $\cb$ is given as
\begin{align}
\probeb{\cb}&=
1-\sum_{n=\tbl{\cb}}^{\tbh{\cb}} \prob{y=n}\nonumber\\
&=
1-\sum_{n=\tbl{\cb}}^{\tbh{\cb}} 
\mathbb{E}_{\phim}\left[
\prob{y=n\mid \phim}
\right] \nonumber\\
&=1-\sum_{n=\tbl{\cb}}^{\tbh{\cb}} 
\mathbb{E}_{\phim}\left[
\frac1{n!}\exp\left(-V(\rd,\cb,\phim)\right)\right.\left.\times V(\rd,\cb,\phim)^n
\right] ,\label{eq:AppC1}
\end{align}
\newcommand{\laplace}[2]{\mathcal{L}_{#1}\left(  #2\right)}
where the last step is due to \eqref{eq:AppCymdist}. Now, note that, 
\begin{align}
&\expU{-Z}(- Z)^n=
\left.
\frac
{\mathrm{d}^n\exp\left ( - \rho Z \right)}
{\mathrm{d}\rho^n}\right\vert_{\rho=1} \implies\expect{Z^n\expU{-Z}}=
{(-1)}^n
\left.
\frac{	
	\dd^n \laplace{Z}{\rho}
}
{
	\dd\rho^n
}
\right\vert_{\rho=1}.
\end{align}
Applying this identity in \eqref{eq:AppC1}, we get
\begin{align}
\probeb{\cb}
&=1-\sum_{n=\tbl{\cb}}^{\tbh{\cb}} 
\frac1{n!}{(-1)}^n
\left.
\frac{	
	\dd^n \laplace{V(\rd,\cb,\phim)}{\rho}
}
{
	\dd\rho^n
}
\right\vert_{\rho=1},
\label{eq31}
\end{align}
with the slight abuse of notation that $\dfracnrho{n}{F}=F$ for $n=0$. 
Here, $ \mathcal{L}_{V}(\rho)$ is the Laplace transform of $V$ which can be obtained as,
\begin{align}
\laplace{V(\rd,\cb,\phim)}{\rho}&=\mathbb{E}\left [ \exp\left( -\rho\cb N \cirslot{\rd}{0}
-\rho\sum_{\x\in \phim} v(\norm{\x}) 
\right)\right ]\nonumber\\
&=\exp\left(-\rho \cb N h_{\rd}[0]\right)
\mathbb{E}_{\phim}
\left [
\exp\left( -\rho\sum_{\x\in  \phim } v(\|\x\|) \right)\right ]\nonumber\\
&\stackrel{(a)}{=}
\exp\left (-\rho \cb N h_{\rd}[0]-4\pi \lambda\int_{a}^{\infty}(1-\mathbb{E}_{u_{z}[0]}
\left[
\expU{	-\rho h_{z}[0]u_{z}[0]}
\right])
\ z^2\mathrm{d}z
\right )\nonumber\\
&=\exp\left (-
\rho N\cb h_{\rd}[0]-4\pi \lambda \pone  \int_{a}^{\infty}
\left(
1-\expU{-\rho h_{z}[0]N}
\right)\ \ z^2\dd z\right ),\label{gpe4}
\end{align}
where $(a)$ is due to the marked version of Campbell theorem.
By successive differentiation of  \eqref{gpe4} using Bell polynomial version of Faa di Bruno's formula \cite[eq.(2.2)]{rio}, we get
\begin{align}
\left.\frac{\dd^n \mathcal{L}_{V(\rd,\cb,\phim)}(\rho)}{\dd\rho^n}\right\vert_{\rho=1}&=
{ (-1)}^n
\expS{ - \rho\cb Nh_{\rd}[0]-4\pi \lambda 
	\pone
	\int_{a}^{\infty}(1 -\exp( - h_{z}[0]N ))\ z^2\mathrm{d}z}\nonumber\\
&\times
\mathfrak{B}_{n} (\mathbf{\noisiqfunc}(\rd,\lambda)),\label{eq33}
\end{align}
where $\mathbf{\noisiqfunc}(\rd,\lambda)=[\noisiqfunc_{1}(\rd,\lambda),\noisiqfunc_{2}(\rd,\lambda),\cdots$ $,\noisiqfunc_{\eta-1}(\rd,\lambda)]$ with 
\begin{align}
\noisiqfunc_m(\rd,\lambda)&= N\cb \cirslot{\rd}{0}
\1(m=1)+4\pi \lambda 
\pone \int_{a}^{\infty}\expU{ -h_{z}[0]N} {(h_{z}[0]N )}^m z^2\dd z.
\end{align}

Now, 
\begin{enumerate}
	\item $\probeb{0}$: Substitute \eqref{eq33} in \eqref{eq31} with $\cb=0$, and we get \eqref{eq12}.
	
	\item  $\probeb{1}$: Note that 
\begin{align}
\probeb{1}
&=1-\sum_{n=\eta}^{\infty} 
\frac{{(-1)}^n}{n!}
\left.
\frac{	
	\dd^n \laplace{V(\rd,1,\phim)}{\rho}
}
{
	\dd\rho^n
}
\right\vert_{\rho=1}\nonumber\\
&=\sum_{n=0}^{\eta-1} 
\frac{{(-1)}^n}{n!}
\left.
\frac{	
	\dd^n \laplace{V(\rd,1,\phim)}{\rho}
}
{
	\dd\rho^n
}
\right\vert_{\rho=1}\label{eq:appCpeb1}.
\end{align}
 Substitute \eqref{eq33} in \eqref{eq:appCpeb1} with $\cb=1$, and we get \eqref{eq13}.
\end{enumerate}

\section{Proof of Theorem-2}\label{app:D}

Note that 
$V(\rd,\cb,\phim)$ for $\cb=0$ is independent of $\rd$. Therefore $\probeb{0}$ for the case when $\rd$ is a random variable, would be the same as $\probeb{0}$ in Theorem \ref{thm:fixedrdnoisi}.

$\probeb{1}$ for the case when $\rd$ is random, is given as
\begin{align}
\probeb{1}&=\sum_{n=0}^{\eta-1} 
\mathbb{E}_{\rd,\phim}\left[
\frac1{n!}{\expS{-V(\rd,1,\phim)}{V(\rd,1,\phim)}^n}
\right] \\
&=\sum_{n=0}^{\eta-1} 
\frac{{(-1)}^n}{n!} \expects{\rd}{
	\left.
	\frac{	
		\dd^n  {\laplace{V(\rd,1,\phim)}{\rho}}
	}
	{
		\dd\rho^n
	}
	\right\vert_{\rho=1}}.
\end{align}
Now, from \eqref{eq33}, 

\begin{align}
\probeb{\cb}
&=\sum_{n=0}^{\eta-1} 
\frac{1}{n!}\mathbb{E}_{\rd}\left[
	\exp \left( - \rho Nh_{\rd}[0]-4\pi \lambda 
	\pone
		\int_{a}^{\infty}(1 -\exp( - h_{z}[0]N ))\ z^2\mathrm{d}z\right)
	\mathfrak{B}_{n} (\pmb{\beta}(\rd,\lambda))
\right]\label{eq:AppD1}.
\end{align}
Now recall that
$\pmb{\beta}(\rd, \lambda)=[\alpha_{1}(\lambda)+N
\cir{\ts}{\rd}$ $
,\alpha_{2}(\lambda),\cdots ,\alpha_{\eta-1}(\lambda)]$.  
Using the following property of Bell polynomials
	\begin{align}
	&\mathfrak{B}_n (x_1+y,x_2,x_3,\cdots,x_n)=\sum_{i=0}^{n}\binom{n}{i}
	\mathfrak{B}_{n-i} (x_1,x_2,x_3,\cdots,x_n)(y)^i,
	\end{align}
in \eqref{eq:AppD1} and then, changing the order of expectation and summation, we get \eqref{eq21}.

\section{Proof of Theorem \ref{thm:fixedrdisi}}\label{app:E}

Let us denote the current bit of the tagged TBN as $\cb$.  Hence, $u_{\xd}[0]=\cb N$. 

Let $\w(\norm{\x})$ be the expected number of IMs including ISI that were emitted by the transmitter located  at $\x$ and absorbed by the RBN. {\em i.e.},
\begin{equation}
\w(\norm{\x})=\sum_{l=0}^{L-1}\cirslot{\norm{\x}}{l} u_{\x}[ -l].
\end{equation}
Let $\wC$ be  the expected sum number of received molecules conditioned on $\phim$ {\em i.e.},
\begin{align}
\wC(\rd,\cb,\phim)&=\cb N\cirslot{\rd}{0}+\sum_{l=1}^{L-1}\cirslot{\norm{\xd}}{l} u_{\xd}[ -l]+\sum_{\x\in \phim} \w(\norm{\x}).
\end{align}
Now, given $\phim$, the total number of IMs received is Poisson distributed \ie 
\begin{align}
y\mid \phim &\sim \mathcal{P}(\wC(\rd,\cb,\phim)).\label{eq:AppEymdist}
\end{align}
Similar to the proof of Theorem \ref{thm:fixedrdnoisi}, the probability of incorrect decoding for bit $\cb$ is given as
\begin{align}
\probeb{\cb}
&=1-\sum_{n=\tbl{\cb}}^{\tbh{\cb}} 
\frac{{(-1)}^n}{n!}
\left.
\frac{	
	\dd^n \laplace{\wC(\rd,\cb,\phim)}{\rho}
}
{
	\dd\rho^n
}
\right\vert_{\rho=1}.\label{eq:AppE1}
\end{align}
%
The Laplace transform of $\wC(\rd,\cb,\phim)$ is given as
 \begin{align}
\laplace{\wC}{\rho}
&=\exp\left(-\rho \cirslot{\rd}{0}\cb N \right)\times\mathbb{E}_{u_{\rd}[-(L-1):-1]}
\left[\exp\left(\sum_{l=1}^{L-1}h_{\rd}[l]\ud[-l]\right)\right]\nonumber\\
&\times\mathbb{E}_{\phim}\left [ \exp\left( -\rho\sum_{\x\in  \phim }\w( \norm{\x}) \right)\right ].\label{eq45}
\end{align}
%
%
Here, $\mathbb{E}_{\ud[-(L-1):-1]}$ represents the expectation with respect to number of molecules emitted by the transmitter at $\xd$ in $L-1$ previous time slots before the current slot 0. Now,
\begin{align}
\mathbb{E}_{\ud[-1:-L+1]}\left[\exp\left(\sum_{l=1}^{L-1}-\rho h_{\rd}[l]\ud[-l]\right)\right]&=\prod_{l=1}^{L-1}\mathbb{E}_{\ud[-1:-L+1]}\left[\exp\left(-\rho h_{\rd}[l]\ud[-l]\right)\right]\nonumber\\
&=\prod_{l=1}^{L-1}\left(P_{\mathrm{0}}+P_{\mathrm{1}}\exp\left(-\rho h_{\rd}[l]N\right)\right).\label{eq46}
\end{align}
The last term in \eqref{eq45} can be simplified as
\begin{align}
\mathbb{E}_{\phim}\left [ \exp\left( -\rho\sum_{\x\in  \phim }\w( \norm{\x}) \right)\right ]
&=\exp\left ( -4\pi\lambda\int_{a}^{\infty}\left(1-\right.\right.\nonumber\\
&\left.\left.\mathbb{E}_{u_{z}}\left[\exp
\left( -\rho \sum_{l=0}^{L-1}h_{z}[l]u_{z}[-l] \right)
\right]\right)\ z^2\dd z\right )
\nonumber\\
&=\exp\left( -4\pi\lambda\int_{a}^{\infty}\left(1-\right.\right.\left.\left.\prod_{l=0}^{L-1}\left(\pzero
	+\pone\exp\left(-\rho h_{z}[l]N\right)\right)\right)\ z^2\dd z\right).\label{eq47}
\end{align}
Substituting \eqref{eq46} and \eqref{eq47} in \eqref{eq45} gives,
\begin{align}
\laplace{\wC(\rd,\cb,\phim)}{\rho}&=\prod_{l=1}^{L-1}\left(\pzero +\pone
\exp\left(-\rho h_{\rd}[l]N\right)\right) \times\exp\left (-\rho h_{\rd}[0]N \cb  -\right.\nonumber\\
&\left.4\pi\lambda\int_{a}^{\infty}\left[1-\prod_{l=0}^{L-1}
\left(\pzero+\pone\expU{-\rho h_{z}[l]N}
\right)\right]\ z^2\dd z\right ).\label{gpe26}
\end{align}
To evaluate \eqref{eq:AppE1}, we need to find the $n\ths$ derivative of $\laplace{\wC}{\rho}$. 
Now let us define the following $L$ functions $F_1(),\cdots F_L()$:
\begin{align}
&F_\ell(\rho)=\left(\pzero +\pone
\exp\left(-\rho h_{\rd}[\ell]N\right)\right) \text{ for } \ 1\le \ell \le L-1, \text{and}\nonumber\\
&F_L(\rho) =\exp\left (-\rho h_{\rd}[0]N \cb  -4\pi\lambda\int_{a}^{\infty}\left[1-\prod_{l=0}^{L-1}
\left(\pzero+\pone\expU{-\rho h_{z}[l]N}
\right)\right]\ z^2\dd z\right ),\label{eq:FLDef}
\end{align}
such that
\begin{align}
\laplace{\wC(\rd,\cb,\phim)}{\rho}=&F_1(\rho)F_2(\rho)\cdots F_L(\rho).
\end{align}
Using the General Leibniz rule \cite{olver2012applications} in the above equation, we get
\begin{align}
&\dfracnrho{n}{
	\laplace{\wC(\rd,\cb,\phim)}{\rho}
}=
\sum_{ m_1+...+ m_L=n}\frac{n!}{ m_1! \cdots m_L!}\prod_{1\leq \ell \leq L}
\dfracnrho{m_\ell}{F_\ell},\label{eq:laplacewderivative}
\end{align}
%
%
where the sum extends over all $L$-tuples $( m_1,..., m_L)$ of non-negative integers with ${\sum _{m=1}^{L} m_{m}=n}$. 
Note that for $m=0$, the $m\ths$ derivative $\dfracnrho{m}{F(\rho)}=F(\rho)$. \\
The $m\ths$ derivative of $F_\ell(\rho)$'s can be computed as
\begin{align}
\dfracnrho{m}{F_\ell(\rho)}& =\pzero \1( m=0)+
\pone {(-h_{\rd}[\ell]N)}^{ m}\exp\left(-\rho h_{\rd}[\ell] N\right)\nonumber\\
&={(-h_{\rd}[\ell]N)}^{ m} \left(\pzero \1( m=0)\right.\left.+\pone \expS{-\rho h_{\rd}[\ell] N}\right),\label{eq:Fellderivative}
\end{align}
for $0\le \ell\le L-1$.

To calculate the $m\ths$ derivative of $F_L(\rho)$, we will use Bell polynomial version of Faa di Bruno lemma to get
\begin{align}
\dfracnrho{m}{F_L(\rho)}&=F_L(\rho) \mathfrak{B}_{ m} (\pmb{\isiqfunc}(\rd,\lambda)),\label{eq:FLderivative}
\end{align}
with $\pmb{\isiqfunc}(\rd,\lambda)=[\isiqfunc_{1}(\rd,\lambda),\isiqfunc_{2}(\rd,\lambda),\cdots$ $,\isiqfunc_{ \eta-1}(\rd,\lambda)]$ . Here, $\isiqfunc_i$ denotes the $i\ths$ derivative of the exponent term in \eqref{eq:FLDef} and is given as
\begin{align}
&\isiqfunc_i(\rd,\lambda)= - \cirslot{\rd}{0}N \cb \indside{i=1}+4\pi \lambda \int_a^\infty \dfracnrho{i}{}\left(
\prod_{l=0}^{L-1}
\left(\pzero+\pone\expU{-\rho h_{z}[l]N}
\right)
\right) z^2 \dd z.\label{eq:qivalue}
\end{align}
Now, the derivative terms in the second term can be computed using General Leibniz rule as
\begin{align}
\dfracnrho{i}{}\left(
\prod_{l=0}^{L-1}
\left(\pzero+\pone\expU{-\rho h_{z}[l]N}
\right)
\right)
&=\sum_{q_0+\cdots q_{L-1}=i}
\frac{i!}{q_0!\cdots q_{L-1}!}
\prod_{l=0}^{L-1} \left(
\pzero \indside{q_l=0}\right.\nonumber\\
&\left.+\pone {(-1)}^{q_l}
{(\cirslot{z}{l}N)}^{q_l}\expU{-\rho N \cirslot{z}{l} } 
\right)\nonumber\\
&=\sum_{q_0+\cdots q_{L-1}=i}
\frac{i! {(-1)}^iN^i}{q_0!\cdots q_{L-1}!}
\prod_{l=0}^{L-1} \left(
{\cirslot{z}{l}}^{q_l}
\right)\left(
\pzero \indside{q_l=0}+\pone\expU{-\rho N \cirslot{z}{l} } 
\right).\label{eq:qivalue2}
\end{align}

Now, substituting \eqref{eq:qivalue2} in \eqref{eq:qivalue} and then, substituting the resultant value in \eqref{eq:FLderivative} with some arrangement of terms, we get,
\begin{align}
\dfracnrho{m}{F_L(\rho)}&={(-1)}^m F_L(\rho) \mathfrak{B}_{ m} (\pmb{\isiqfunc'}(\rd,\lambda)),\label{eq:FLderivativetwo}
\end{align}
with
$\pmb{\isiqfunc'}(\rd,\lambda)=[\isiqfunc'_{1}(\rd,\lambda),\isiqfunc'_{2}(\rd,\lambda),\cdots,$ $\isiqfunc'_{\eta-1}(\rd,\lambda)]$  and 
\begin{align}
\isiqfunc'_i(\rd,\lambda)&= \cirslot{\rd}{0}N \cb \indside{i=1}+4\pi \lambda \int_a^\infty 
\sum_{q_0+\cdots q_{L-1}=i}
\frac{i! N^i}{q_0!\cdots q_{L-1}!}
\prod_{l=0}^{L-1} \left(
{\cirslot{z}{l}}^{q_l}
\right)\nonumber\\
&\times\left(
\pzero \indside{q_l=0}+\pone\expU{-\rho N \cirslot{z}{l} } 
\right)
z^2 \dd z.
\end{align}

Substituting the values from \eqref{eq:Fellderivative} and \eqref{eq:FLderivativetwo} in \eqref{eq:laplacewderivative}, we get,
\begin{align}
\dfracnrho{n}{
	\laplace{\wC(\rd,\cb,\phim)}{\rho}
}
&=
\sum_{ m_1+...+ m_L=n}\frac{n!{(-1)}^n}{ m_1! \cdots m_L!}\prod_{\ell=1}^{L-1}
{(h_{\rd}[\ell]N)}^{ m_\ell} \left(\pzero \1( m_\ell=0)\right.\nonumber\\
&\left.+\pone \expS{-\rho h_{\rd}[\ell] N}\right)\times  F_L(\rho) \mathfrak{B}_{ m_L} (\pmb{\isiqfunc'}(\rd,\lambda))\label{eq:laplaceWderivativefinal}.
\end{align}
Now, 
\begin{enumerate}
	\item $\probeb{0}$: Substitute \eqref{eq:laplaceWderivativefinal} in \eqref{eq:AppE1} with $\cb=0$, and we get \eqref{eq24}.
	
	\item  $\probeb{1}$: Note that 
		\begin{align}
	\probeb{1}
	&=1-\sum_{n=\eta}^{\infty} 
	\frac{{(-1)}^n}{n!}
	\left.
	\frac{	
		\dd^n \laplace{\wC(\rd,1,\phim)}{\rho}
	}
	{
		\dd\rho^n
	}
	\right\vert_{\rho=1}=\sum_{n=0}^{\eta-1} 
	\frac{{(-1)}^n}{n!}
	\left.
	\frac{	
		\dd^n \laplace{\wC(\rd,1,\phim)}{\rho}
	}
	{
		\dd\rho^n
	}
	\right\vert_{\rho=1}\label{eq:appEpeb1}.
	\end{align}
		Substitute \eqref{eq:laplaceWderivativefinal} in \eqref{eq:appEpeb1} with $\cb=1$, and we get \eqref{eq25}.
\end{enumerate}



\ifCLASSOPTIONcaptionsoff
  \newpage
\fi

\bibliographystyle{IEEEtran}

\end{document}